\providecommand{\tightlist}{%
  \setlength{\itemsep}{0pt}\setlength{\parskip}{0pt}}

\DeclareUnicodeCharacter{0301}{\hspace{-1ex}\'{ }}

\PassOptionsToPackage{table,xcdraw,dvipsnames,prologue}{xcolor}

\documentclass[sigplan, nonacm]{acmart}

\usepackage{mdframed}
\usepackage[utf8]{inputenc}
\usepackage{multirow}
\usepackage{tikz}
\usetikzlibrary{backgrounds}
\usetikzlibrary{positioning}
\usepackage{tcolorbox}
\usepackage{threeparttable}
\AtBeginEnvironment{tcolorbox}{\small}
\usepackage{algorithm}
\usepackage[noend]{algpseudocode}
\usepackage{graphicx}
\usepackage{subcaption}
\usepackage{mwe}
\usepackage{tablefootnote}
\usepackage[table,xcdraw,dvipsnames]{xcolor}
\usepackage[normalem]{ulem}
\useunder{\uline}{\ul}{}
\usepackage{longtable}
\makeatletter
\def\BState{\State\hskip-\ALG@thistlm}
\makeatother

\definecolor{softbrown}{HTML}{3E1F16}
\definecolor{softblue}{HTML}{113F76}
\definecolor{softred}{HTML}{A51C4D}
\definecolor{softteal}{HTML}{006F7F}

\definecolor{purple}{HTML}{9D00D2}
\newtcolorbox{mybox}[3][]
{
  colframe = #2,
  colback  = #2!0,
  coltitle = #2!0!white,  
  title    = {#3},
  #1,
}

\newcommand{\highlighted}[3][yellow]{%
  \tikz[baseline=(word.base)]{%
    \node[inner sep=1pt] (word) {%
      \textcolor{#2}{#3}%
    };
    \begin{pgfonlayer}{background}
      \fill[#1, fill opacity=0.3, rounded corners=3pt]
        ([xshift=-2pt, yshift=-2pt]word.south west) rectangle
        ([xshift=2pt, yshift=2pt]word.north east);
    \end{pgfonlayer}
  }%
}

\newcommand{\username}[1]{%
  \textcolor{softblue}{%
    \raisebox{-0.5ex}{\includegraphics[height=1em]{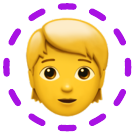}}~\textsf{\small#1}%
  }%
}

\def\checkmark{\tikz\fill[scale=0.4](0,.35) -- (.25,0) -- (1,.7) -- (.25,.15) -- cycle;} 
\AtBeginDocument{%
  }

\copyrightyear{2025}
\begin{document}

\title[If You Had to Pitch Your Ideal Software]{If You Had to Pitch Your Ideal Software --- Evaluating Large Language Models to Support User Scenario Writing for User Experience Experts and Laypersons}


\author{Patrick Stadler}
\affiliation{%
  \institution{HU Berlin}
  \city{Berlin}
  \country{Germany}
}
\orcid{0000-0002-5224-1522}
\email{patrick.stadler@charite.de}

\author{Christopher Lazik}
\affiliation{%
  \institution{HU Berlin}
  \city{Berlin}
  \country{Germany}
}
\orcid{0000-0002-8687-8548}
\email{christopher.lazik@hu-berlin.de}

\author{Christopher Katins}
\affiliation{%
  \institution{HU Berlin}
  \city{Berlin}
  \country{Germany}
}
\orcid{0000-0001-6257-7057}
\email{christopher.katins@hu-berlin.de}

\author{Thomas Kosch}
\affiliation{%
  \institution{HU Berlin}
  \city{Berlin}
  \country{Germany}
}
\orcid{0000-0001-6300-9035}
\email{thomas.kosch@hu-berlin.de}

\renewcommand{\shortauthors}{Stadler et al.}

\begin{abstract}

The process of requirements analysis requires an understanding of the end users of a system. Thus, expert stakeholders, such as User Experience (UX) designers, usually create various descriptions containing information about the users and their possible needs. In our paper, we investigate to what extent UX novices are able to write such descriptions into user scenarios. We conducted a user study with 60 participants consisting of 30 UX experts and 30 novices who were asked to write a user scenario with or without the help of an LLM-supported writing assistant. Our findings show that LLMs empower laypersons to write reasonable user scenarios and provide first-hand insights for requirements analysis that are comparable to UX experts in terms of structure and clarity, while especially excelling at audience-orientation. We present our qualitative and quantitative findings, including user scenario anatomies, potential influences, and differences in the way participants approached the task.
\end{abstract}




\maketitle


\section{Introduction}\label{introduction}
The central principle of human-centered design is to place the user at the heart of the design process, enabling designers to gain insight into their needs and incorporate this understanding into interactive systems \cite{holtzblatt1993making}. An effective approach to understanding and identifying user needs involves extracting possible requirements, followed by constructing formal abstractions. This can be achieved by capturing these requirements for instance in user scenarios and use cases. User scenarios are narrative descriptions of a user's perspective regarding a certain system, including their role, needs, and challenges, while use cases are formal representations of a user's interaction with a system \cite{kulak2012use} often presented as small units described in a special technical notation (e.g. Unified Modeling Language, UML) or in short text sections. Writing user scenarios and use cases for requirement analysis and, ultimately, system design involves various stakeholders at different stages with the idea of a generalizable perspective on specific features the potential users could want or challenges that they could experience, respectively. 
In an optimal case, designers and other stakeholders find and interview the right people (e.g., potential users or customers) to translate their knowledge into user scenarios while keeping them concise and unambiguous \cite{robeer2016automated}.


However, many artifacts of requirements analysis, including user scenarios, are often perceived as rather artificial tools as an end to support user research -- replicating some \emph{outsider}'s perspective on a system's actual \emph{users}. Ultimately, potential users of a novel application or interactive systems are usually only included rather late in the requirements analysis phase, resulting in applications that are not adaptive or tailored to the individual user's requirements in the long run, such as changed requirements or different conditions at the end of the research. But what if we assigned the development of user scenarios to the actual end users of an application with the idea of crafting software that actually matches the respective user's intent? A potential problem is that users who are not proficient in writing user scenarios do not have the necessary proficiency and experience to provide high-quality scenarios. 

In this context, Large Language Models (LLMs) are postulated to support the user-centered design by supporting scenario writing~\cite{fang2023systematic} or requirements solicitation~\cite{ferrari2024model, ferrari2024model1}. LLMs can assist users, including laypersons, in crafting user scenarios to better align software with user intent. Although LLMs have demonstrated impressive results in understanding language backed by Natural Language Processing (NLP) \cite{robeer2016automated, sinhaText2TestAutomatedInspection2010}, the sudden and rapid rise of LLMs also opens up entirely new possibilities to empower laypersons in formulating their own user scenarios. With the help of LLMs, laypersons can, together with user experience experts, contribute to more precise user scenarios. 

Although LLMs show great promise in supporting writing, there remains a research gap in investigating the quality of written user scenarios. While numerous studies evaluate the use of LLMs in requirements analysis \cite{ferrari2024model}, they usually focus on the stakeholders' perspectives rather than a system's potential users' views. 

We believe that the individual hand-written perspective of laypeople could truly enrich the existing approaches found in requirements analysis, as it could capture unique, first-hand user insights that experts or fully automated systems may overlook. Supporting the laypeople without taking away their own ideas and writing style is different from a fully automated system, where the user's perspective can also be determined, e.g. by an LLM automatically writing a user scenario based on certain data points, but may not be able to accurately capture the finer details.

Our study investigates whether and to what extent laypersons can write user scenarios using an LLM-supported writing assistant. It also examines how effectively off-the-shelf LLMs can interpret these scenarios to generate formal and technical use cases understandable by adaptive systems, serving as artifacts that can be processed by both, humans and further (AI-based) systems. We examined how well laypersons compare to UX experts when writing user scenarios by qualitatively assessing the scenarios and resulting use cases through two UX experts. We conducted a user study with 60 participants in a 2\,$\times$\,3 factorial design divided into two groups (i.e., UX experts and laypersons) and three LLM configurations (i.e., no, zero-shot, and few-shot writing assistant, refer to Section \ref{methodology}) to learn

\begin{enumerate}
\def\labelenumi{\alph{enumi})}
\tightlist
\item
  whether laypersons could produce meaningful user scenarios and use cases without and with the help of an AI writing assistant and LLMs
\item
  how challenging the process was, and what problems and opportunities occur
\item
  how suitable the generated use cases are for an automated system adaption
\end{enumerate}

Our \textbf{contributions} are the following:

\begin{itemize}
\tightlist
\item
  a: Design and implementation of an LLM-supported writing assistant, including prompt instructions.
\item
  b: An investigation of the quality of user scenarios created by both UX experts and non-experts by either using or not using a writing assistant.
\item
  c: Design considerations for user scenario writing and use case generation with LLM-supported writing assistants.
\end{itemize}

\section{Related Work}\label{related-work}

\subsection{Use of NLP and LLMs in Requirements Analysis}\label{use-of-nlp-and-llms-in-requirements-analysis}
The use of NLP and LLMs in requirements analysis has a great spark in recent HCI research due to the new possibilities of LLMs. Various studies examined the application of such methods for user story automation, use case generation, diagram generation from user stories or use cases, creating abstractions of user stories, or simply extracting certain elements from user stories and use cases \cite{raharjana2021user}. Research applications cover a wide range of the potential or involved stakeholders, the individual tasks where LLMs are used, or which processes they try to automate, such as from a text-only level to user interface design. According to experts in agile software development, two of the key challenges are finding and writing down user requirements from an end user's point of view and changing requirements, for example, by varying demands or simply over time \cite{schon2017key}.
Current research shows great potential for LLMs to be used in requirements analysis, for instance, by supporting or improving various formal descriptions and artifacts, such as enhancing user stories \cite{zhang2024llm}, and use cases \cite{de2023echo}, to evaluate the completeness and accuracy of software requirements \cite{white2024chatgpt, ronanki2023investigating}, and to support UX designers in their design process \cite{ekvall2023integrating}. In addition, the use of LLMs in requirements analysis is applied to construct various artifacts, such as graphical user interfaces \cite{kolthoff2024interlinking}.
Besides potentially enhancing processes like writing requirements, these LLMs can assist in automating tasks, such as generating test cases from user stories \cite{sami2024tool}. Another example found in recent research utilized ChatGPT to generate Unified Modeling Language diagrams from requirement documents \cite{ferrari2024model1}. Although the results were promising, the authors found multiple challenges and issues, many of which are related to specific attributes of large language models, such as terminology (made-up terms, inconsistent terms, and misunderstood terms), lack of contextual understanding, and variants of produced output. The lack of context was particularly significant when different requirements were linked to each other, which also posed a challenge in terms of the size of the context window.
To address the issue of an incomplete perspective, for example, when leaving out non-experts, or simply to streamline the overall process, researchers proposed and evaluated different approaches, such as visualizing user stories \cite{jahan2024automated, herwanto2024leveraging, gilson2020generating, elallaoui2018automatic}. Using this technique, the authors aimed to visually represent core concepts and processes that were initially written in textual user stories, making it easier for different types of stakeholders to understand the information. An interesting aspect of \cite{gilson2020generating} is the way the authors connected different user stories that follow a similar structure into a \emph{bigger picture}.

\subsection{User Scenarios and Use Cases}
\label{sec:user_scenarios_and_use_cases}
User scenarios describe the individual user's perspectives on a virtual novel interactive system \cite{kc2024analysis}, that not only include a rather brief expectation but also an extensive description of a user's daily work routine, the people they collaborate with, as well as the use or misuse of technology. These user scenarios are a type of user story. However, they are different from the \emph{\textless As a user, I want x to obtain y\textgreater{}} notation as proposed by Cohn \cite{cohn2004user} in that these user scenarios are a broader description and contain much more detail. Thus they are also referred to as complex stories \cite{welinuser}. User scenarios are narrative stories about people and their activities \cite{carrol1999five} written in natural language and are written from the perspective of a user's daily life experiences with a system \cite{liu2012scenario}. Ultimately, user scenarios are an effective tool to be used in the requirement analysis process, as they provide a much more detailed view of the actual user needs \cite{damas2005generating}.

User scenarios usually aim to capture an authentic view regarding the interaction or possible interaction with a (software) system from a user's perspective. Thus, user scenarios are usually structured around the goals or possible improvements (or resolutions), the current workflow or routine, one or more problem statements, and challenges or pain points that the user experienced \cite{lucassen2017extracting, michailidou2013create, van2013exploring, the-interaction-design-foundation-2024}. 

For use cases, there are many definitions of what they could or should be, what information they carry, etc \cite{kulak2012use, tiwari2015systematic}. A straightforward definition is that of Cockburn \cite{cockburn1997structuring}:
``A use case is a collection of possible sequences of interactions between the system under discussion and its external actors, related to a particular goal''. Ultimately, use cases are part of the requirements analysis \cite{kulak2012use} and can be seen as formal documentation of user needs in sequential and logical order. 

\subsection{Involving Laypersons in Requirements Analysis}\label{involving-non-technical-stakeholders}

Although LLMs arrived and manifested quickly in computer science and HCI research, and chat-like interfaces such as ChatGPT allow non-technical users to use this technology conveniently; there still are many fields and applications hidden behind a conceptional \emph{barriers} to be accessible by the general public, such as understanding, coordination, and information \cite{ko2004six}. Including laypersons offers a great opportunity if this technology is prepared and made accessible accordingly. Regarding user experience, involving non-technical stakeholders opens the process of requirements analysis to a broader public, and both stakeholders, as system designers and engineers, and users could benefit from the exchange of different perspectives, leading to a more holistic and realistic perspective. When it comes to AI technology, there are some known pitfalls but also great opportunities \cite{yang2018grounding}. Researchers tried different approaches to involve laypersons in the process, such as no-code interfaces \cite{zamfirescu2023johnny}. One possible way to involve users in the requirement analysis process is by outsourcing the process of authoring user stories or user scenarios to users \cite{kc2024analysis}. LLMs can help extract use case components from these user-authored user scenarios, while these LLMs benefit from additional domain knowledge and it helps to improve the resulting quality of the generated use case components, such as exact definitions, for instance, of the concept of a \emph{step} in use case descriptions.
Another approach is to utilize user-generated data that already exist, such as reviews for automated requirements analysis, for instance, from app stores \cite{ghosh2024exploring}, or automatically extracting sub-features of app descriptions \cite{wei2024getting}. 

\subsection{Supporting Laypersons With AI Writing Assistance}\label{ai-writing-assistance}
AI-based writing tools are designed to assist users in crafting narratives, such as stories and more formal texts. These writing assistants vary widely in their capabilities, ranging from basic corrective technologies like spell checkers intended to enhance current text to more advanced support for creative processes such as story development or idea generation \cite{fang2023systematic}.
Previous research showed that the perception of such writing assistants is influenced by factors such as individual preferences, context, and the type of suggestions they give \cite{wester2024exploring}.

Writing requirements can be seen as both a creative task, as it requires the LLM to come up with their own ideas that are not necessarily given by the user, and a form of augmentation, as they may help struggling users expand their own abilities or digital competencies \cite{wester2024exploring}. Common questions in HCI and AI research revolve around whether LLMs can be \emph{creative} in any sense. This is important for the machine-based analysis of user stories and the development of suitable outcomes, especially when trying to use LLMs for creative technology exploration, as LLMs need to ``think'' outside the box to come up with ideas for interaction techniques based on the user stories. Current research indicates that some LLMs can actually write creatively to some extent \cite{gomez2023confederacy}. Nevertheless, a significant challenge persists: we are unable to accurately forecast or manage the outcomes of these LLMs. Their auto-regressive characteristics further limit their ability to deliver entirely creative ideas \cite{franceschelli2023creativity}. Thus, many researchers examined possible applications where the writing assistant mostly supports the users and enhances their own abilities to write, for instance, in creative writing, as proposed by \cite{tost2024futuring}, or scientific writing \cite{gero2022sparks}. 

Another application is a writing assistant for multimodal creative writing \cite{singh2023hide}. The authors used distinct datasets to fine-tune their models, such as a movie summaries corpus and a writing prompts dataset. An interesting question found in recent research regarding such writing assistants is whether users actually incorporate the suggestions into their process or if there are concerns or restraints. The authors showed that some of their participants did not use the suggestions at all for a variety of reasons, one of which was that they felt it encouraged a stereotypical writing style, while others mentioned that the writing assistant distracted them.

\section{Methodology}\label{methodology}
We conducted a user study with both UX experts and laypersons who were asked to write a custom user scenario with or without LLM support based on the condition. The writing assistant is available in two variants: zero-shot and few-shot. In the zero-shot condition, the LLM was guided solely through task-specific instructions, while in the few-shot condition, the prompt included example scenarios alongside the instructions \cite{kojima2022large}. Next, two experienced researchers from our lab independently rated all user scenarios. We additionally provide in-depth analyses of various user scenarios and describe common patterns, and aspects of well and poorly written user scenarios.

\vspace{2mm}
\noindent We present the following three research questions:

  $\mathbf{RQ_{1}}$: Are laypersons able to write reasonable user scenarios with and without AI-backed writing assistants?

  $\mathbf{RQ_{2}}$: Are there differences between zero-shot and few-shot approaches?

  $\mathbf{RQ_{3}}$: How useful are LLM-generated use case previews in user scenario writing?

\section{Study}\label{sec:study}
We conducted a user study where UX experts and UX laypersons had to write user scenarios with or without an LLM-supported writing assistant. Participants could also automatically generate use case previews from their user scenario anytime as another artifact to examine how their scenario compiles to a more formal description. We explored how both user groups approached the process of writing user scenarios, their methodology, and whether and to what extent the user scenarios ultimately differed.

\subsection{Participants}\label{participants-1}

We hired 60 participants via Prolific with various backgrounds for our between-subjects study. 
The UX experts (9 self-identified as female) ranged from 20 to 65 years (\(\overline{Age}\) = 33.7 years, SD = 13.47 years).
The laypersons (10 self-identified as female) ranged from 22 to 67 years (\(\overline{Age}\) = 33.8 years, SD = 11.44 years).

For the UX experts, we applied the following screener sets: participants with knowledge of UX and software development techniques, with a function in engineering (e.g. software), product or product management, project or program management, or research, an approval rate of 99-100\footnote{The approval rate is an indicator based on the total number of rejections and approvals for a participant's submissions}, and a minimum number of previous submissions of 1000\footnote{Selecting a rather high number of previous submissions helps to select reliable participants}.

For the UX laypersons, we applied several criteria. We targeted individuals based on their frequency of device usage, focusing on those who used their devices at least once a week, up to multiple times per day. Participants were required to have at least a basic understanding of computers. We further narrowed our selection to individuals who were employed full-time or part-time and whose work involved significant interaction with computers or mobile devices. Specifically, participants were included if they used such devices for at least 25\% of their work. This approach ensured that the selected participants experienced using technology in their daily lives.

To prevent low-quality submissions and cheating through external AI, such as ChatGPT, we have implemented several measures: maintaining an approval rate of 99-100\%, requiring at least 1000 previous submissions, explicitly stating that using external AI will result in rejection and reporting, requiring participants to confirm via a checkbox at the study's start that AI use is prohibited, logging all system interactions, and analyzing written user scenarios for common AI-generated artifacts. Participants were also invited to share an optional screen recording while doing the experiment for an additional bonus of \$5.00. Ultimately, eleven participants have sent us their screen recordings.

Despite our efforts and indications, we had to reject and resample eight submissions (3 laypersons and 5 UX experts) due to low quality or use of external AI.

\paragraph{Post-Study Survey}
We asked the participants to answer some brief questions after they completed the study, including their previous experience with AI technology and their level of experience with writing or reading user scenarios, as well as a free text option to provide additional feedback on the writing assistant. 

The average self-reported confidence with AI was 4.2 (SD=$1.19$) for the UX experts and 3.8 (SD=$0.85$) for the laypersons on a Likert scale ranging from $1..5$ (no confidence..high confidence). 
For the UX experts, nine participants reported 1-2 years of AI experience, 13 reported 3-5 years, four had more than 5 years, and four had less than one year or no experience at all. A total of 20 laypersons noted having 1-2 years of AI experience, 6 reported 3-5 years, and four indicated less than one year or no experience with AI at all. UX experts rated their competence in writing or interpreting user scenarios on an average of 4.13 (SD=$1.00$) on a Likert scale ranging from $1.5$ (no competence..high competence).

\subsection{Procedure}\label{procedure-1}
We implemented a 2\,$\times$\,3 factorial design (refer to Table \ref{tab:study_design}). We randomized the conditions to ensure a balanced distribution among all participants. 
The sessions lasted around 30 minutes (UX experts) and 34 minutes (laypersons) on average. Participants were paid a compensation of \$14/hour for their time. The idea was to compare the quality of user scenarios between typical stakeholders in requirement analysis and a sample of random users who had no experience with the topic, in general, to examine whether laypersons can produce quality user scenarios that can be translated into use cases for a formal abstraction. To prepare participants for the tasks, we shared a brief presentation and introduction to the study objective and provided a brief user scenario example to help them get started.

\begin{table}[h]
  \centering
  \caption{The 2\,$\times$\,3 factorial design with the conditions expertise (experts vs. laypersons) and writing assistant. The baseline is the experts without writing assistant condition because it will be a replicate of the status quo, i.e., the traditional process of writing user scenarios and use cases. For the writing assistant (WA) option, we used three conditions: no writing assistant, with zero-shot prompting, and with few-shot prompting (includes the data collected in section 4). †denotes the minimal assistance without suggestions on how to write or improve the user scenario.}
  \label{tab:study_design}
  \resizebox{.4\textwidth}{!}{%
    \begin{tabular}{lcc}
      \toprule
      & \textbf{Experts} & \textbf{Laypersons} \\
      \midrule
      \textbf{Without WA\small\dag} & Baseline & A \\
      \textbf{Zero-Shot WA} & B & C \\
      \textbf{Few-Shot WA} & D & E \\
      \bottomrule
    \end{tabular}%
  }
\end{table}

For the baseline and A conditions, only minimal assistance was presented to the participants. This assistant only gave simple suggestions, such as using the preview or complete buttons or regularly saving the project, as these functions were also present in all other conditions where the actual writing assistant could be used by the participants. We wanted to offer these steps to all participants, regardless of whether they had the writing assistant at reach or not, as previewing the generated use cases could make some difference in this experiment. The option of also regularly checking the completeness of the user scenario was also important, as it gave some indication of the general progress. For the baseline and A condition, no additional hints or information were shown to the participants; in all other conditions, such additional information was shown in a tooltip when hovering over the complete button, for example, about how to complete the user scenario.

\subsection{Task}
Both groups, the UX experts and the laypersons, received specific instructions addressing the distinct viewpoints on the task, which are summarized below. 

\paragraph{UX Experts}
UX experts were asked to write a user scenario for a third person, for example, a colleague, a partner, or a friend. We chose this approach to prevent the later revealing of user scenarios that were written by UX experts, which would probably spoil the ratings and reviews by independent researchers. In addition, we wanted to get a more realistic scenario for the experts as they would usually also write such user scenarios on other people than themselves. Also, this allowed use to compare the both approaches more realistically, especially regarding how genuine and authentic the expert views are aligned with real user requirements. 

\paragraph{Laypersons}
Non-expert participants were asked to write a user scenario for their profession or occasion from their own perspective as if they were to pitch it to a team of software engineers. They were asked to share a complete view, including their current workflow, motivation, interactions, and collaboration with co-workers. In addition, they should add any challenges or pain points with their current setup, such as bottlenecks and inefficient or inconvenient tasks within their workplace. 

\paragraph{Steps}
In addition to the instructions, all participants were required to log in to the web app using the username provided via Prolific and their Prolific ID, consent to the informed consent form, and agree to the no-AI statement to proceed with the study. Participants opened their projects and accessed the text editor to write down the user scenarios.

The writing assistant, located on the bottom left of the screen, could be used at any time during the process. Participants could preview automatically generated use cases based on their user scenario by using the preview button. Once finished, participants clicked the "End" button in the top left corner, triggering a popup with further instructions and a short survey. After completing the survey, they received the final code for Prolific, which is used as a confirmation token that a participant has successfully completed the study.

Participants were also encouraged to share any screen and audio recordings they made to help us better understand their thought processes and workflow.

\subsection{The Probe}\label{widgetexplorer-llm-studio}

\begin{figure*}[t]
\centering
\includegraphics[width=\textwidth]{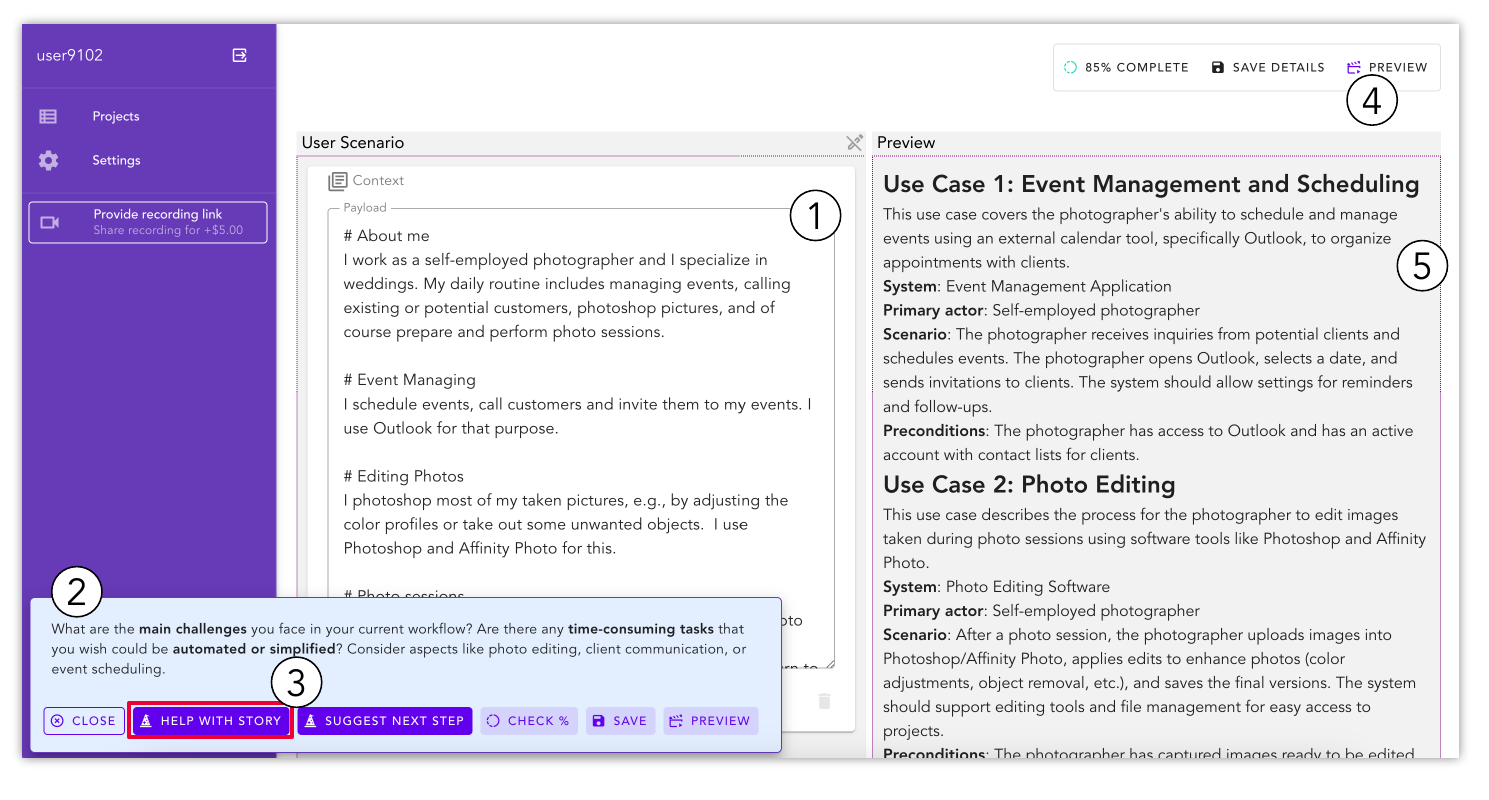}
\caption{Screenshot of the web interface used to conduct the main study. Participants were asked to write down their user scenarios \raisebox{.5pt}{\textcircled{\raisebox{-.8pt} {1}}}. The LLM-based writing assistant \raisebox{.5pt}{\textcircled{\raisebox{-.8pt} {2}}} helps users with two different LLM-supported options \raisebox{.5pt}{\textcircled{\raisebox{-.8pt} {3}}}, including improving the current user scenario, and suggesting a next step. Note, that the button in the red rectangle is only presented in the full writing assistant, not the minimal version (baseline and A conditions). 
The user scenario can then be sent to the LLM middleware using the preview button \raisebox{.5pt}{\textcircled{\raisebox{-.8pt} {4}}} which returns the generated use cases of an LLM for reference \raisebox{.5pt}{\textcircled{\raisebox{-.8pt} {5}}}. \label{fig:widgetexplorer_screenshot}}
\Description{Screenshot of the probe for the main user study, showing the interface including the user scenario editor, the use case preview, the writing assistant, and important buttons such as preview.}
\end{figure*}

We designed and implemented a custom frontend as well as the corresponding services which was used as the probe for our study. With the help of this tool, we had an easy-to-setup experiment software that allowed us to use the LLMs and the writing assistant, as well as a built-in database. Previewing the generated use cases was not mandatory to write the user scenarios, but it allowed the participants to verify their input at any time and test out how their user scenarios were processed accordingly. A screenshot of the used probe can be seen in figure \ref{fig:widgetexplorer_screenshot}.

\subsection{LLMs and Instructions}
\label{llms-and-instructions}
We used two different LLMs throughout our study, where one model was used for the writing assistant, and the other model was used for generating the use case previews. We used the basic LLMs as provided through the official APIs, without fine-tuning. For the few-shot writing assistant (conditions D and E) we collected various examples of user scenarios for each specific task (writing assistance and use case generation) as prompting examples, which were appended to each call to the respective service. The few-shot examples are described in Section \ref{fewshot-examples}. To reduce the output variance and make the results more predictable, we set all temperatures to the default setting (\(1.0\)), and used defaults for all other parameters. We provided system instructions for both the writing assistant and the use case generation, which we also share in this section. 
Table \ref{tab:llms} overviews all used LLMs with their settings. All LLMs are commercial closed-source models. We decided to use GPT-4 and Claude because of their accessibility and availability and for better comparison, as they were often used in previous HCI research \cite{kevian2024capabilities, hochmair2024correctness, gupta2024evaluation} and because they were more reliable than other commercial models, such as Gemini, especially regarding following instructions and the requested formatting. After a pretest with five participants who wrote several user scenarios and generated use cases, we selected Claude for the writing assistant and GPT-4 for the use case generation task, as these models seemed to work best for the respective tasks.

\begin{table}[]
\centering
\begin{threeparttable}
\caption{Overview of the used LLMs in our study. One was used for the writing assistant suggestions, the other for the use case (UC) generation.}
\label{tab:llms}
\begin{tabular}{@{}llll@{}}
\toprule
\textbf{LLM}    & \textbf{Suggestions} & \textbf{UC Preview} & \textbf{Version} \\ 
\midrule
\textbf{GPT-4}   &                      & \checkmark                  & gpt-4o-mini\tnote{a} \\
\textbf{Claude}  & \checkmark           &                             & claude-3-5-sonnet\tnote{b} \\
\bottomrule
\end{tabular}
\begin{tablenotes}
\footnotesize
\item[a] \url{https://openai.com/index/gpt-4o-mini-advancing-cost-efficient-intelligence/}
\item[b] \url{https://www.anthropic.com/news/claude-3-5-sonnet}
\end{tablenotes}
\end{threeparttable}
\end{table}

\subsubsection*{Prompts and System Instructions}\label{prompts-and-system-instructions}

\label{sec:system_instructions}
To craft the prompts we consulted several sources. A good starting point was the paper by White et al. \cite{white2024chatgpt}, which provides interesting insights and prompt templates for multiple challenges found in requirements analysis, such as requirements elicitation and refactoring. To craft the individual prompts for the use case preview and the writing assistant, we mainly drew on three sources: existing user scenarios available online and from a preceded data collection (refer to Section \ref{fewshot-examples}), best practices from LLM guides (such as the OpenAI\footnote{\url{https://platform.openai.com/docs/guides/prompt-engineering}} and Anthropic\footnote{\url{https://docs.anthropic.com/en/docs/welcome}} docs), and prior research regarding requirements analysis tasks. There are many strategies and guidelines available to optimize the instructions for LLMs, such as providing context \cite{sorensen2022information}, or using personality modifiers \cite{wei2024leveraging}. We applied several of these guidelines to our pre-prompts and tested their appliance and outcome before the study. 

\begin{figure*}[t]
\includegraphics[width=.85\textwidth]{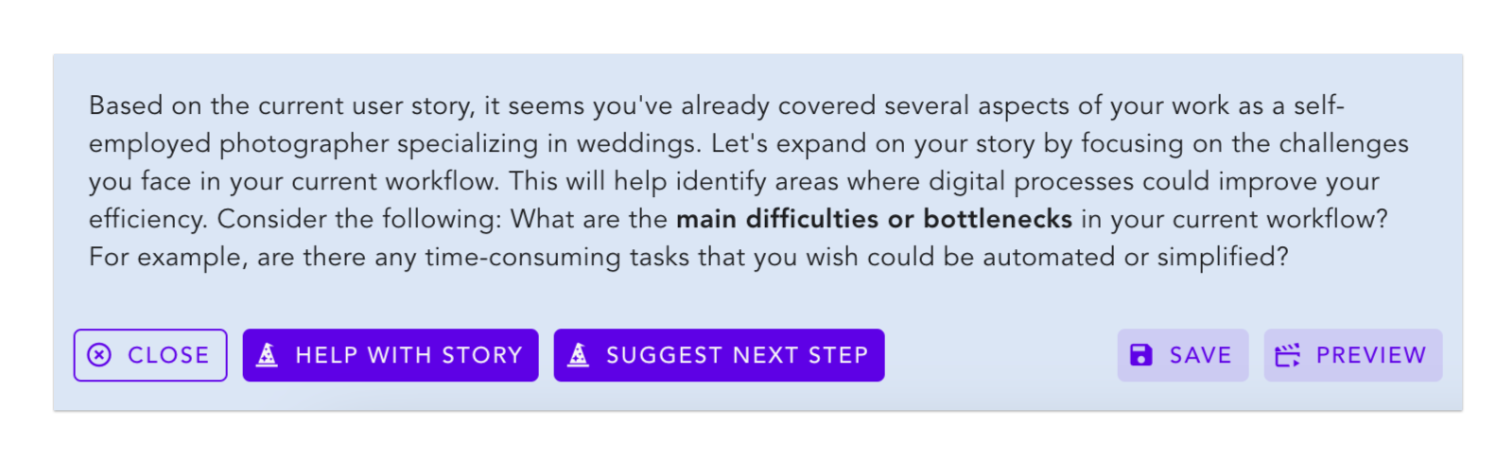}
\caption{Close-Up of the LLM-backed writing assistant, as presented to the participants, showing a suggestion that asks to describe the main bottlenecks\label{fig:wa_closeup}}
\Description{Screenshot of the writing assistant close-up, showing a suggestion and the buttons for Help with the story, suggest next step, check completeness, save, and preview.}
\end{figure*}

For the user scenario assistance, we included the common elements of user scenarios as described in Section \ref{sec:user_scenarios_and_use_cases}. We also added the two aspects collaboration and technology, as we were also interested to see how participants interacted with others (e.g., co-workers), and the technology they used. 

\paragraph{\textbf{Few-Shot Training Examples}}
\label{fewshot-examples}
For a complete overview of all conditions and the use of few-shot examples, refer to Table \ref{tab:conditions_wa}.

\textbf{User Scenarios:} To prepare examples for the few-shot writing assistant, we curated various user scenarios from available online sources\footnote{https://nfdi4culture.de/resources/user-stories.html, accessed 12/2024}\textsuperscript{,}\footnote{https://www.justinmind.com/blog/how-to-design-user-scenarios/, accessed 10/2024}. 
We used these user scenarios as few-shot training examples for the writing assistant in the study (see Section \ref{sec:study}), similar to the one listed in Section \ref{sec:appendix_user_scenario}. 



We then included ten of these annotated pairs into the system prompt which will only be used for the few-shot conditions (D and E).

\textbf{Use Cases:}
For the use case generation, we also prepared several use cases from various online sources\footnote{https://www.wrike.com/blog/what-is-a-use-case/, accessed 11/2024} and added them to the prompt, equivalent to the user scenarios. The few-shot examples for the use cases were utilized regardless of the writing assistant's condition, meaning they were applied across all conditions (A$..$E).

\paragraph{\textbf{Use Case Preview}}
To formulate concrete steps, we consulted several details found in prior research. For example, the findings from \cite{kc2024analysis} contain descriptions of extracting specific components from a user scenario, such as actors, goals, or necessary steps. The system prompt comprises two segments: a pre-prompt containing the role, a brief task description and context, and an example section that consists of several use cases as described in Section \ref{tab:uc_system_prompt}.
To generate the use case previews for the user scenarios, we instructed GPT-4 to carefully read the provided user scenario, identify and extract the main aspects of the main actors (such as the user of the scenario), and break down the user scenario into discrete activities or interactions. It was then tasked with writing concise use cases containing a title, one or more actors, a description, and the preconditions. 

\paragraph{\textbf{Writing Assistant Suggestions}}
We collected various examples of user scenarios as few-shot examples. These examples were comprised of a short task description and the user scenario, and one or more matching annotations and reviews from our own data collection, resulting in multiple example pairs.

To introduce a role and context, we started to provide some introduction about the LLM's personality, the task, and what should be covered in the user scenario.
We also included the instructions we gave the participants as additional context for working on the task, some guidelines, and principles of good user scenarios extracted from the collected data (refer to Section \ref{fewshot-examples}). Together with further instructions regarding formatting and the length of the suggestions, the writing assistant returned brief suggestions (less than 300 characters per invocation) to ensure that the participants would still need to think about the suggested aspects on their own.
Refer to Table \ref{tab:wa_comp_zs_fs} for some example suggestions in a direct comparison between the zero-shot and few-shot prompts.

\subsection{LLM-based Writing Assistant}
\label{llm-based-writing-assistant}
This section will outline several internal components of the LLM-backed writing assistant. The writing assistant was available to all participants regardless of their assigned condition. However, all LLM-supported features were only available in the zero-shot and few-shot conditions (B, C, D, E) and not in the baseline and A conditions. 

\begin{table*}[t]
    \centering
    \caption{Summary of the functionality across all six conditions of the writing assistant (WA). All conditions but baseline and A had LLM-supported writing assistance, i.e., suggestions that were generated using an LLM. The baseline and A conditions only had a minimal assistant capable of previewing the generated use cases and saving the current user scenario. \dag = only completeness percentage without further information. \textsuperscript{1}Few-Shot examples used in writing assistant. \textsuperscript{2}Few-Shot examples used in use case generation.}
    \label{tab:conditions_wa}
    \resizebox{\textwidth}{!}{%
    \begin{tabular}{lccccccccc}
        \toprule
        \textbf{Condition} & \textbf{Assistance (WA)} & \textbf{LLM-Support} & \textbf{Help with story} & \textbf{Suggest next step} & \textbf{Completeness} & \textbf{Preview} & \textbf{Save} & \textbf{FS WA\textsuperscript{1}} & \textbf{FS UC\textsuperscript{2}} \\
        \midrule
        \textbf{Baseline, A} & $\cdot$ & $\cdot$ & $\cdot$ & $\cdot$ & minimal$\dag$ & \checkmark & \checkmark & $\cdot$ & \checkmark \\
        \textbf{B, C} & Zero-Shot WA & \checkmark & \checkmark & \checkmark & \checkmark & \checkmark & \checkmark & $\cdot$ & \checkmark \\
        \textbf{D, E} & Few-Shot WA & \checkmark & \checkmark & \checkmark & \checkmark & \checkmark & \checkmark & \checkmark & \checkmark \\
        \bottomrule
    \end{tabular}%
    }
\end{table*}

The writing assistant was designed as a small overlay in the bottom left corner of the editor screen. To ensure that all participants knew the writing assistant, it automatically appeared when the editor was opened and drew attention to itself by flashing for a moment. Besides the LLM-supported features (Help with the story, Suggest next step), the writing assistant featured three buttons to check completeness, save the current user scenario, and generate the use case previews.

\subsubsection*{Available Suggestions}
The LLM-backed writing assistant featured three main suggestion types which are described below. For conditions baseline and A, only the second button (Suggest a Next Step) was shown that only made suggestions containing no LLM features (e.g., reminding a participant to click the preview button).

\paragraph{\textbf{Help With Story}}
Whenever a participant requests this suggestion, the writing assistant analyzes the provided (intermediate) user scenario and suggests how to start, continue, or improve it. Some possible suggestions to help with the beginning of the user scenario writing include questions on the basic idea of what they work, what their profession is called, or what a typical workday looks like. The more the user scenario develops, the more advanced suggestions are given, for instance, what the main responsibilities and primary tasks are, which tools and technologies they use, and with whom they collaborate. To complete the user scenario, suggestions include mentioning challenges with their current workflow and goals. 
The full prompt for this function is listed in Table \ref{tab:wa_help_story}.

\paragraph{\textbf{Suggest a Next Step}}
Participants also had the option to suggest a random next step from a list, such as refining unclear sentences or expressions, reminding users to reflect on their technology usage and preferences in their current work setting, analyzing the provided user scenario based on the pre-prompted metrics and offering suggestions, testing the user scenario outcome using the 'Preview' button, saving their work regularly, or clicking the 'Complete' button to check progress and completeness, which also provides indicators and directions for completing the user scenario. Note that, apart from the preview and save steps; all the previously mentioned steps were unavailable in the non-LLM version of the writing assistant (conditions baseline and A). The full prompt for this function is listed in Table \ref{tab:wa_next_step}.

\paragraph{\textbf{Check Completeness}}
This suggestion analyzes the current user scenario and returns an estimated completeness level in percent, as well as a brief summary of possible improvements.

\section{Findings}\label{results}
We present our findings in this section and assign them directly to the corresponding research questions.

\subsection{Assessment of User Scenarios}
\label{sec:assessment_methods}
The primary research question was whether UX laypersons are able to write reasonable and meaningful user scenarios.

We evaluated the user scenarios based on two methods, which will be applied throughout this section:

\vspace*{3mm}

\textbf{a) Expert rating}
To examine the quality of the user scenarios, we used the checklist from \cite{michailidou2015evaluate} to rate and review all user scenarios. Two raters from our lab, who are experienced in user-centered methods and user scenarios, independently rated each user scenario according to the checklist.

\textbf{b) In-depth analysis}
To assess the similarities and differences, as well as the strengths and weaknesses of the user scenarios, we coded all user scenarios based on role, goals, routine, problem statement, challenges / pain points, and improvement based on the common key aspects found in literature, as described in Section \ref{sec:user_scenarios_and_use_cases}. 

\subsubsection*{\textbf{User Scenario Rating}}
The checklist for the ratings comprises five dimensions: content, expression, structure, audience orientation, and effect. Content included assessments of whether the user scenario relied on authentic data, if the characters and situations were typical, and if it was specific and concrete. The expression category evaluated the language used by the character and its clarity. The structure encompassed several elements concerning the facts presented and the user scenario's length. Audience-Orientation focuses on the context given, ensuring an external individual's comprehension. Lastly, the effect category included three elements assessing if the user scenario was memorable and stimulated new ideas (for a complete overview of all rating items, refer to Section \ref{tab:rating_items_us}). 

To evaluate the LLM-generated use cases, we crafted a checklist based on \cite{lilly1999use} and \cite{anda2001quality} that featured the item's plausibility, readability, structure, the involved actors, content, level of detail, realism, consistency, and whether the use cases were correctly identified.

In total, 24 items for the user scenarios, and ten items for the use cases (refer to Tables \ref{tab:rating_items_us}, \ref{tab:rating_items_uc}) were rated on a scale from $0..100$. 
To determine the inter-rater reliability between the two independent raters, we calculated Krippendorff's Alpha \cite{krippendorff2011computing}. Given the result of $\alpha=0.797$, we infer that the level of agreement is acceptable to satisfactory.


\begin{figure*}[t]
\centering
\begin{subfigure}{.49\textwidth}
  \centering
  \includegraphics[width=\linewidth]{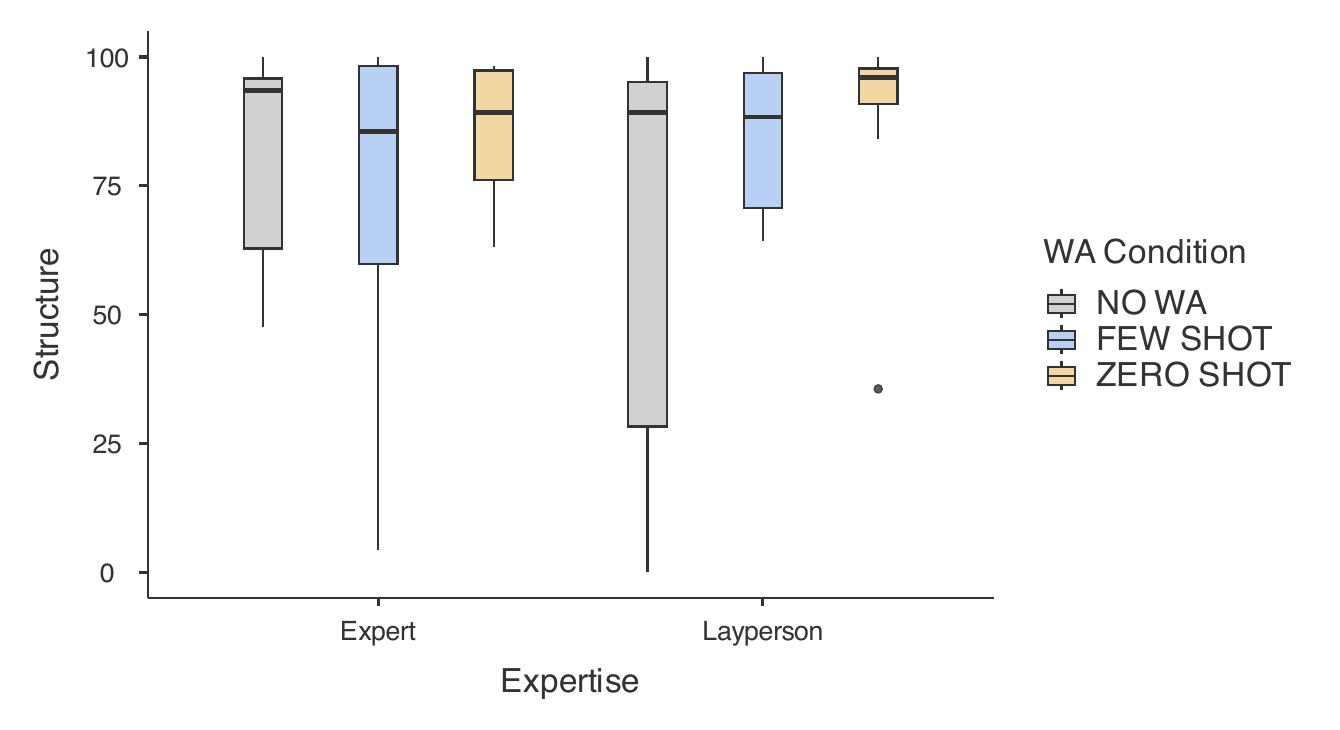}
  \caption{Structure}
\end{subfigure}%
\begin{subfigure}{.49\textwidth}
  \centering
  \includegraphics[width=\linewidth]{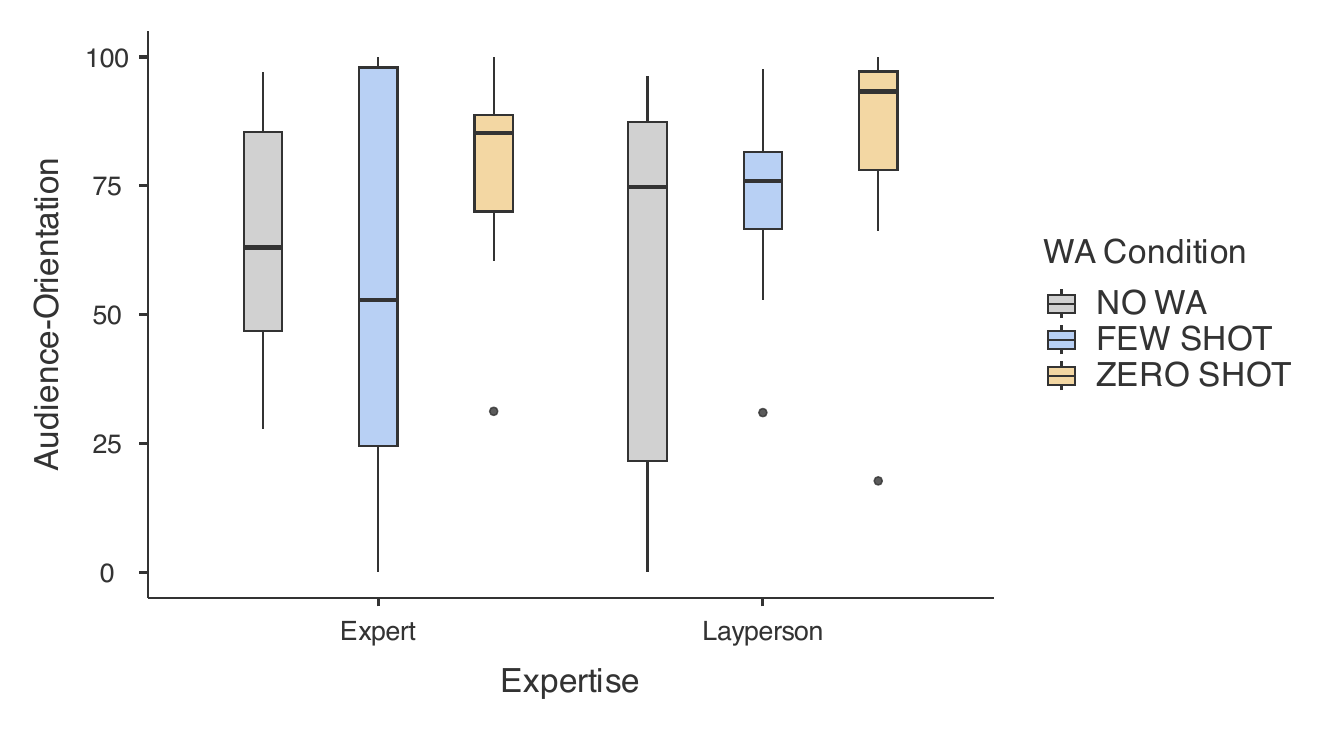}
  \caption{Audience-Orientation}
\end{subfigure}

\bigskip
\begin{subfigure}{.49\textwidth}
  \centering
  \includegraphics[width=\linewidth]{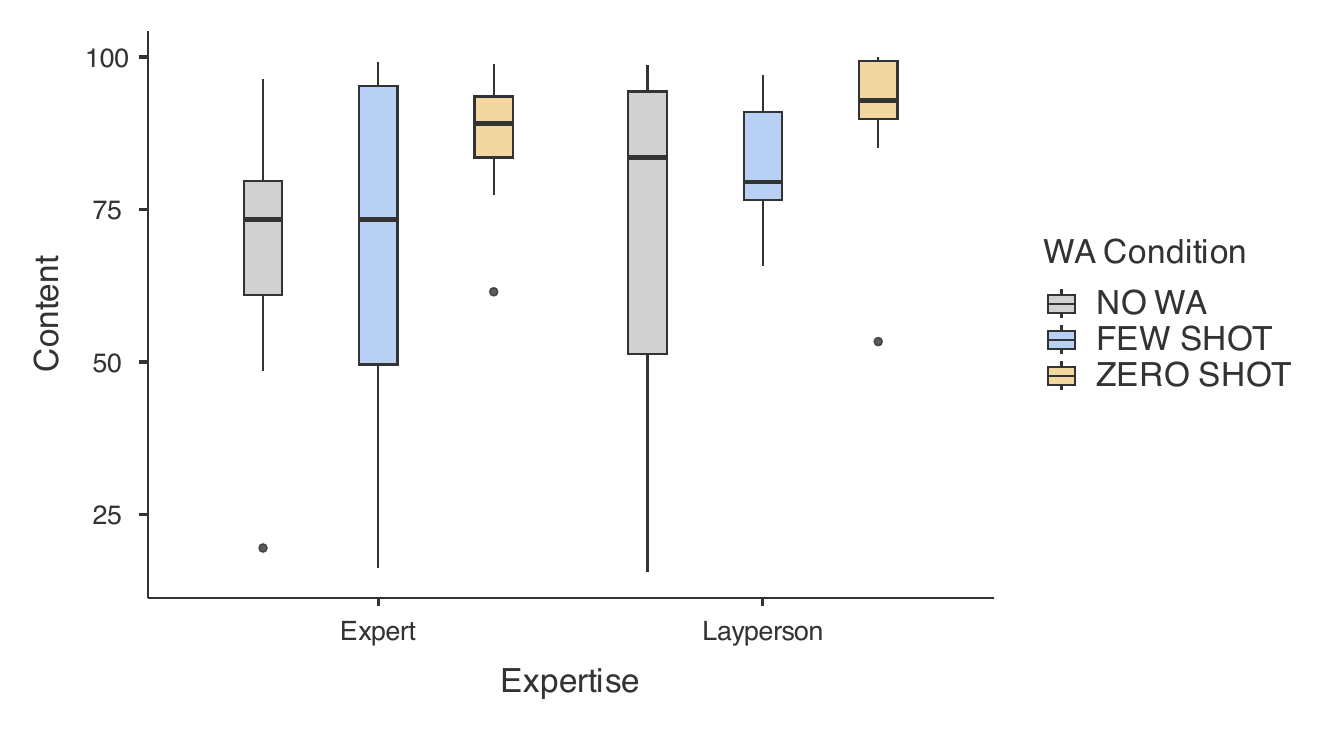}
  \caption{Content}
\end{subfigure}
\begin{subfigure}{.49\textwidth}
  \centering
  \includegraphics[width=\linewidth]{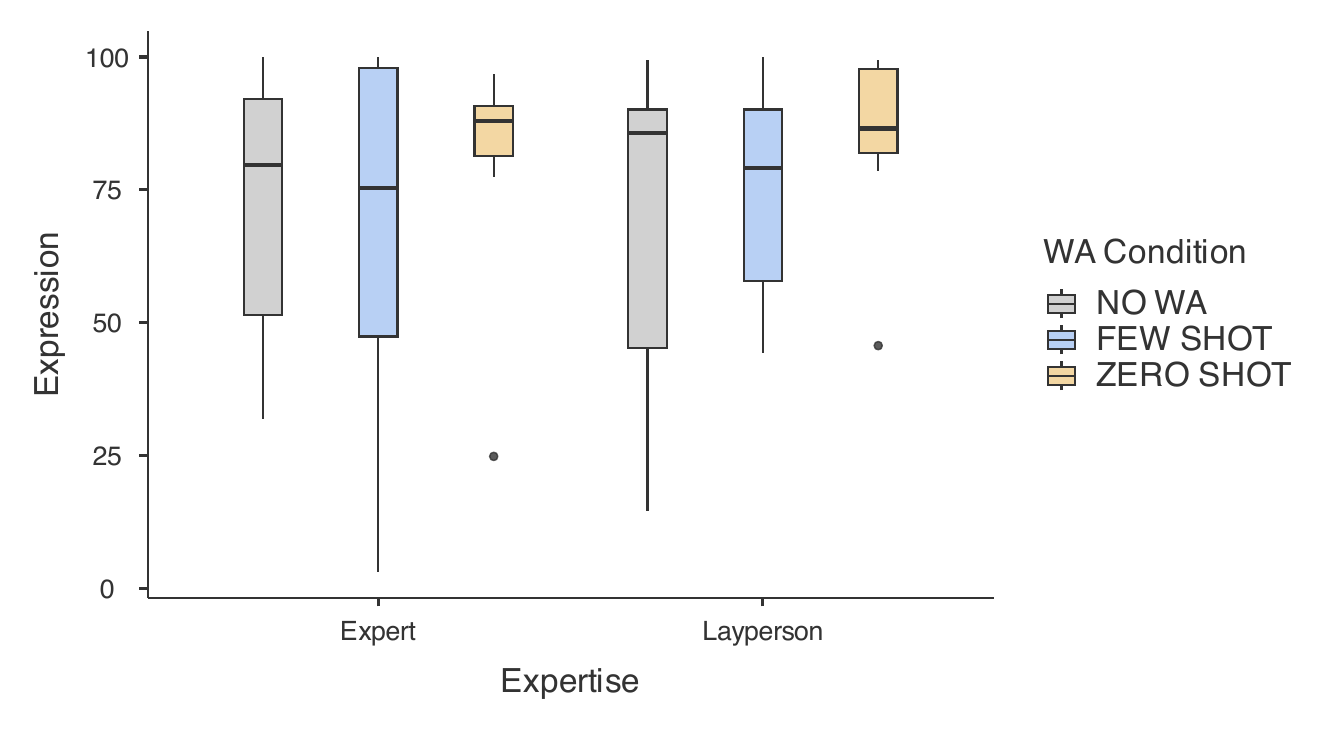}
  \caption{Expression}
\end{subfigure}

\bigskip
\begin{subfigure}{.49\textwidth}
  \centering
  \includegraphics[width=\linewidth]{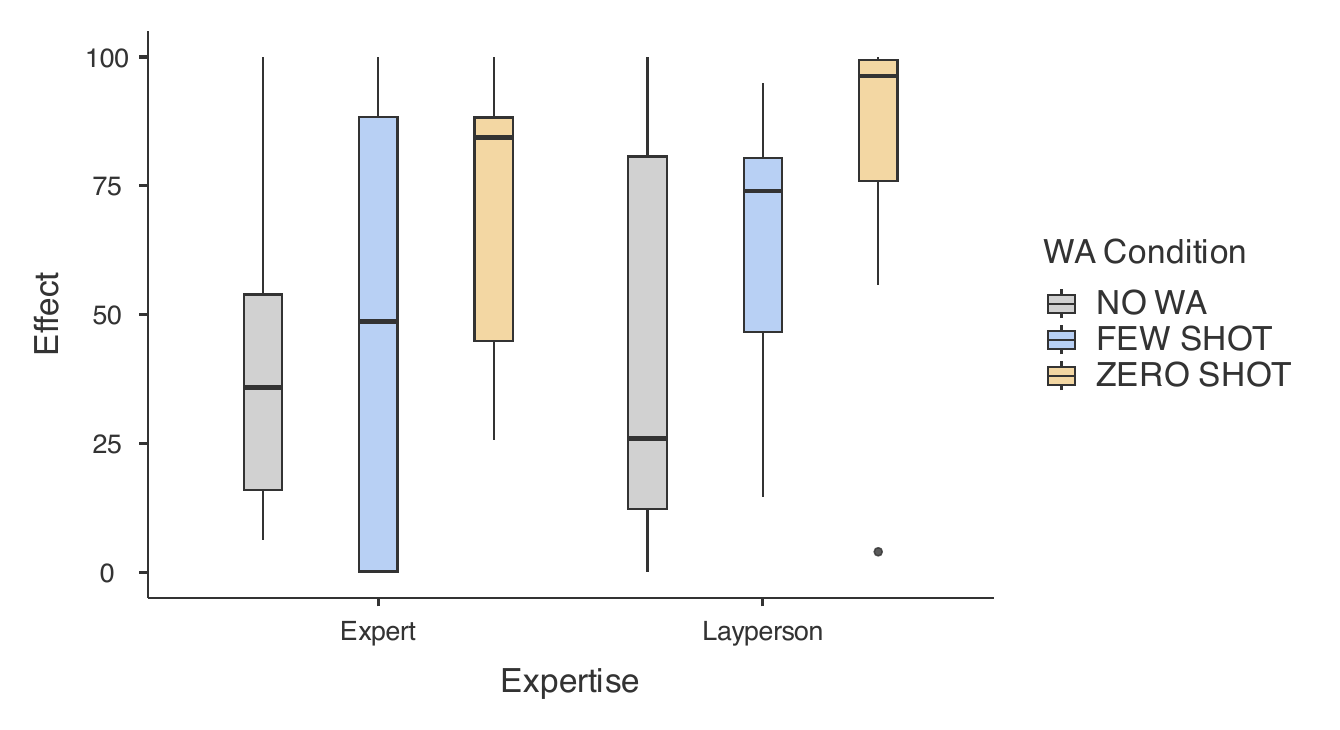}
  \caption{Effect}
\end{subfigure}
\begin{subfigure}{.49\textwidth}
  \centering
  \includegraphics[width=\linewidth]{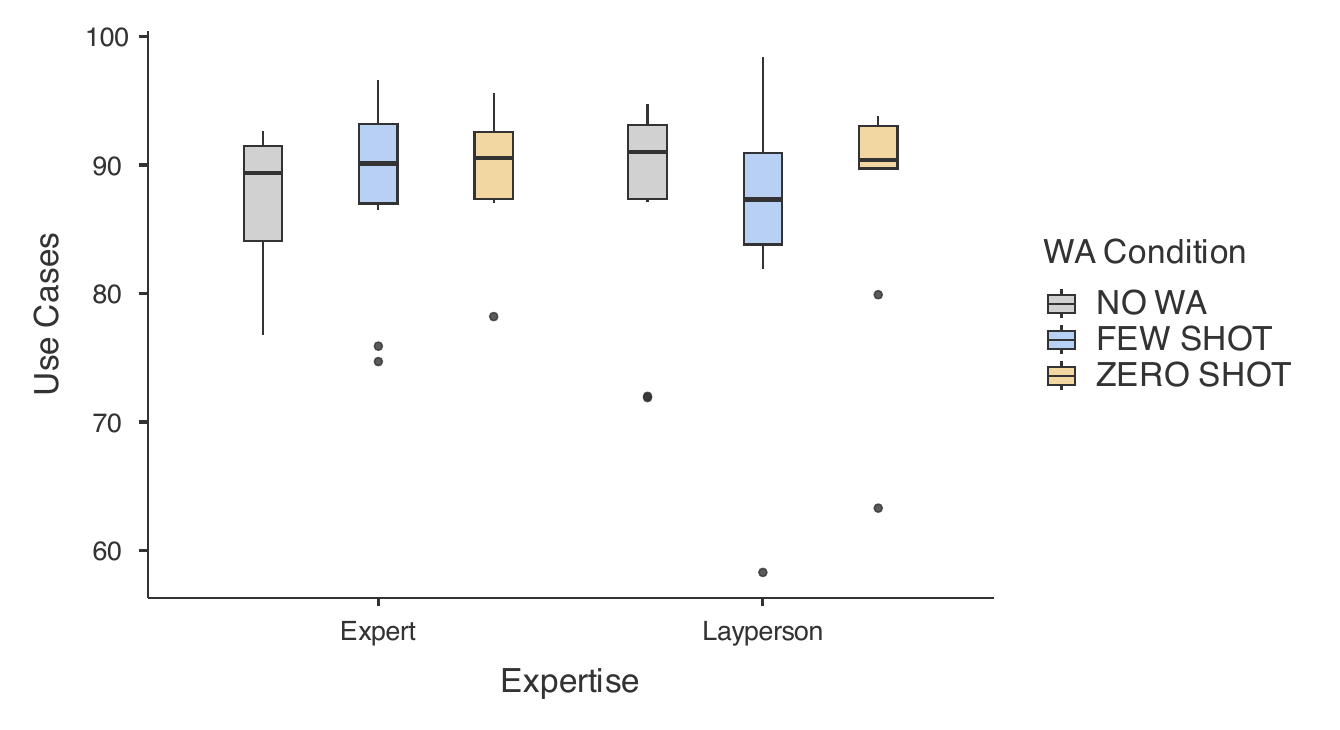}
  \caption{Use Cases}
\end{subfigure}

\caption{Direct comparison of all user scenario ratings per condition (no writing assistant \textit{NO WA}, zero-shot writing assistant \textit{ZERO SHOT} and few-shot writing assistant \textit{FEW SHOT}), and category, as well as the ratings of the generated use cases for reference.}
\label{fig:boxplots}
\Description{TODO}
\end{figure*}

Table \ref{tab:user_ratings} shows the ten best-rated user scenarios. The results show that both UX experts and laypersons can equally write effective and meaningful user scenarios. Eight out of those ten user scenarios were written with the help of the LLM-supported writing assistant. The only exceptions were \username{drummervolcanic}, a UX expert with more than five years of experience, and \username{canmorecontract}, a layperson describing their daily routine as a family restaurant manager. 
To extend the perspective, from the 25 best-rated user scenarios, only four were written without the writing assistant (2x UX experts, 2x laypersons). In addition, 14 of the best 25 user scenarios were written by laypersons (2x without LLM support). 

\begin{table}
    \caption{Authors of the best rated user scenarios. 
    Asterisks (*) indicate that this user scenario is described in-detail in this section. The rating represents the average of $R_{1}$ and $R_{2}$.}
    \label{tab:user_ratings}
    \resizebox{\columnwidth}{!}{%
    \begin{tabular}{ccccc}
        \toprule
        Pos & Participant & Expertise & Condition & Rating  \\
        \midrule
        1 & ebonyballoon* & UX Expert & D & 99.44 \\
        2 & mumbaifoundation & Layperson & C & 98.06 \\
        3 & baghdadpayment* & Layperson & C & 97.84 \\
        4 & helenareview* & Layperson & C & 97.13 \\
        5 & playerspectacle & UX Expert & B & 97.00 \\
        6 & silvermag* & UX Expert & D & 96.02 \\
        7 & drummervolcanic & UX Expert & Baseline & 96.02 \\
        8 & lancasteremergency* & Layperson & C & 95.07 \\
        9 & zippyfrankincense & UX Expert & D & 94.56 \\
        10 & canmorecontract & Layperson & A & 94.28 \\
        \bottomrule
    \end{tabular}
    }%
\end{table}


\begin{table}
    \caption{Authors of the worst rated user scenarios. 
    Asterisks (*) indicate that this user scenario is described in-detail in this section. The rating represents the average of $R_{1}$ and $R_{2}$.}
    \label{tab:user_ratings2}
    \resizebox{\columnwidth}{!}{%
    \begin{tabular}{ccccc}
        \toprule
        Pos & Participant & Expertise & Condition & Rating \\
        \midrule
        52 & aerooctopus & UX Expert & Baseline & 46.60 \\
        53 & frostysullen & UX Expert & D & 39.70 \\
        54 & denimprojector & UX Expert & Baseline & 38.96 \\
        55 & osakaguest & Layperson & C & 38.93 \\
        56 & attendcooing & Layperson & A & 33.26 \\
        57 & amphoradepth* & Layperson & A & 28.36 \\
        58 & beerscrimson* & UX Expert & D & 26.80 \\
        59 & autopsyforsaken & Layperson & A & 24.42 \\
        60 & tophatsunglow & UX Expert & D & 22.10 \\
        \bottomrule
    \end{tabular}
    }%
\end{table}

\subsubsection*{\textbf{User Scenario Analysis}}
After rating all user scenarios, we applied the codes as described in \ref{sec:assessment_methods} to find common patterns, key aspects, and issues in the user scenarios.
An aspect that particularly made these user scenarios good was an extensive and authentic role introduction leading to a meaningful persona. The authors could describe a comprehensible perspective, provide insights into daily life, and provide challenges and pain points. Readers can easily get a picture of the person, as their role is clearly defined and includes relevant aspects that are important to that person. The best-rated user scenarios usually contain a large section on daily routine, the use of technology, and emerging challenges. Compared to other user scenarios, these scenarios' biggest share is the role and routine description. Instead of suggesting specific solutions for a certain challenge, these scenarios offer a detailed insight into the pain points and suggest what could help to improve or overcome those challenges. 

In order to compensate for individual strengths and weaknesses, the extent to which the writing assistant should and may intervene must be assessed. In our study, suggestions were only given if the participants actively requested them. In addition, all suggestions were optional, i.e. the participants had to decide for themselves whether and to what extent they wanted to incorporate them into their user scenario and how they would actually implement the suggestions. Leaving enough room for individuality and output variance is a balancing act that system designers need to be aware of when designing such assistants. 

\subsubsection*{Audience-Orientation}
As an example, the user scenario by \username{baghdadpayment} (layperson, with zero-shot writing assistant) starts with an extensive role description, the daily routine, and an immediate but specific challenge the author had. They propose ways (goals) how to overcome that challenge while leaving enough room for the latter concrete solutions:

\begin{quote}
{\sffamily
    ``\highlighted[softbrown]{black}{Role} \textcolor{softbrown}{My work is in scientific research, specifically quantum chemistry and molecular simulations. This means my work is all computational and I have to interact with several different software platforms to do the functions of my job.} \highlighted[softred]{black}{Challenges} \textcolor{softred}{One of the big challenges is keeping up with which piece of software flows naturally from one to another, in particular with standardizing and properly formatting the types of files I use. The biggest culprit of these are called CIF files (.cif), crystallographic information files, which basically encode all the information for chemical structures of crystalline materials. CIF formatting changes dramatically between different platforms [..].} \highlighted[softteal]{black}{Goals} \textcolor{softteal}{I would propose a software platform where CIF files or other similar chemical file formats can be uploaded and standardized between software platforms, or even modified depending on which software I want to move the CIF file into}''
}
\end{quote}

This user scenario is easy to understand and relatable. In contrast, a user scenario from \username{attendcooing} (layperson, without LLM-supported writing assistant), rated bottom 10, primarily focused on what the author wanted, without specifying any challenges or pain points or why certain suggestions would improve their situation:

\begin{quote}
{\sffamily
``\highlighted[softbrown]{black}{Role} \textcolor{softbrown}{My job is working as an auto adjuster. I interview clients on how the accident happened, and I advise them of liability and how to get their car repaired. I also advise on how to get a rental car for the repair process.} \highlighted[softteal]{black}{Goals} \textcolor{softteal}{I would like to use the newly designed software called the liability assessment module. It will assist me in determining liability in the case of an accident.}

\textcolor{teal}{I would also like to use the newly designed software called the rental car management system, which can assist me when I enter the customer's information, and where they would like to pick up a rental car}''
}
\end{quote}

Although the user scenario starts with a role description, it is difficult to understand what the authors really had in mind, as they did not introduce any challenges or provide parts of their daily routine or similar. 

User scenarios often failed because they aren’t written for an audience. Without considering how other stakeholders will use the scenario, key details are either missing or buried in vague, unstructured language. We found this to be the most common issue.

For instance, take the user scenario written by \username{amphoradepth} (layperson, no writing assistance). This user scenario completely lacks any specification \textit{what} they are working and \textit{which} software they are using, while also asking for vague improvements, such as ``has no bugs'' or ``make me more productive'' that leave the reader in the dark. While certainly this user scenario shares an authentic perspective, it conveys close to zero information as it neither considers the audience, nor follows a common structure.

\begin{quote}
{\sffamily
``How I currently work. I work from home and currently use my laptop and smartphone for work. Some chances I have is that I'm more productive at home then in person. So I feel a software that is tailored to my needs would be a software that is fully functional and is error free. It runs smoothly and has no bugs or issues. Some challenges that would be faced is how good it can actually run if several people use that same software. Will it be able to handle many users? Another challenge is will it actually make me more productive? These challenges I'm sure will be addressed. Modifications will be made to ensure work productivity is high [...]''
}
\end{quote}

The author appears to be expressing thoughts aloud rather than communicating with a clear intent. There seems to be minimal regard for how individuals like designers, developers, or stakeholders could utilize this to grasp user needs or make design decisions.

A key difference between user scenarios that performed well in the audience orientation and effect categories and those that performed rather poorly was the depth of the context and the persona presented. Audience orientation is all about being able to relate to the user scenario and the persona, knowing enough about context, and interpreting the scenario, while the effect is mostly about whether the scenario is memorable and inspires new ideas. User scenarios with insufficient context or incomprehensible role introductions performed substantially worse for those categories.

A good example of a relatable and audience-oriented user scenario is the one about John Doe, a web developer tasked with developing a new website for the government, written by UX expert \username{silvermag}. From a screen recording from that participant, we observed that they invoked the assistant to further develop the persona that they roughly sketched out beforehand. After describing the basic facts about the person, they elaborated on the assistant's suggestion to further describe the role and daily routine. The participant shared their thoughts and gave detailed insights on how they came up with the initial idea for the persona, which seemed to be based loosely on the participant's previous work history as a developer, thus providing meaningful first-hand insights. The entire writing process was a back-and-forth between thinking aloud and invoking the writing assistant, resulting in a top-down approach where details are added constantly. Ultimately, the created persona is relatable, and given all insights and details added as context, the user scenario appears coherent and comprehensible to the audience.

\begin{quote}
{\sffamily
    ``Joe Smith is a web developer working on a contract for a government agency. He is designing a new website for the government. This website will run a web application that will enable US citizens to register for specific government programs that they are required to register for by US law. The agency has a specific set of guidelines about the web application is supposed to function and how it is supposed to look.''
}
\end{quote}

\subsubsection*{Importance of Perspective}
There are numerous examples where laypeople were empowered to write reasonable user scenarios that follow a solid structure that is similar to the ones written by UX experts. In addition, many of these excel the expert-written user scenarios with their detailed and specific perspective. 

For instance, compare the user scenarios written by a UX expert (\username{ebonyballoon}) and the one by layperson \username{helenareview} (zero-shot writing assistant), a remote STEM teacher. The expert created a relatable persona and scenario: Mary, a first-year college student, who is trying to find an apartment in a new city. The scenario describes a hypothetical app for finding off-campus apartments with agent assistance and additional services, and suggests some key features and goals while leaving the concrete implementation open to the future reader. While the user scenario is comprehensible, it is less specific than the one written by the layperson, especially in terms of the actual challenges and goals. 

\begin{quote}
{\sffamily
    ``\highlighted[softbrown]{black}{Role} \textcolor{softbrown}{Mary is just starting her first year in college and wishes to live off-campus, But she is new in the city and does not know anywhere}, \highlighted[softred]{black}{Challenges} \textcolor{softred}{finding an apartment herself would be stressful, and she might not be able to get what she really wants, hiring a house agent can lead to scams or they might charge exorbitant prices.}
    [..] \highlighted[softteal]{black}{Goals} \textcolor{softteal}{she talks about how the application saved her from a stressful search, how it helped her find a safe area with low crime rate, how it helped her avoid scams with agents, and how it helped her with cleaning and decorating with a little fee.} \highlighted[softbrown]{black}{Role} \textcolor{softbrown}{Mary is not tech savvy, but she knows how to navigate through her iPhone.}''
}
\end{quote}

The challenges and goals in the expert-written user scenario are rather shallow and exchangeable, such as the need to find an apartment with certain properties, while scenario by \username{helenareview} clearly identifies specific challenges and goals that are closely related to the daily routine.

\begin{quote}
{\sffamily
    ``\highlighted[softbrown]{black}{Role} \textcolor{softbrown}{I am in the field of Education, specifically, remote education.} The way that I work can be a bit randomized but I always try to follow a protocol and a step by step workflow in order to minimize my time and maximize my results. \highlighted[JungleGreen]{black}{Routine} \textcolor{JungleGreen}{I start by opening Canvas, I got to the module that I am teaching and look at the content there, I download lessons and materials, then I go to my bookmarks and comb through which edu-tech apps would supplement the lesson}. [...]
    \highlighted[softred]{black}{Challenges} \textcolor{softred}{There are many specific tasks that are time consuming. The grading and transferring takes a significant amount of time, aligning standards to the lessons takes a lot of time, finding supplemental activities for each lesson for each day takes an incredible amount of time, When grading work, comparing a student's work against the rubric takes time, creating the rubric takes time.} [...]
    \highlighted[softteal]{black}{Goals} \textcolor{softteal}{My goals for improvement would be that I could focus more on the creative aspects for planning and development without having to focus so much on the administrative daily tasks that are required for each student. I'd like to streamline larger task into manageable steps}'' 
}
\end{quote}

\subsubsection*{Stereotypes and Bias} We observed that some of the user scenarios contained exaggerated traits of the sketched personas and roles, such as the reproduction of stereotypes. These issues along with subjective bias in user scenarios and personas has been examined extensively in prior research \cite{turner2011stereotyping, marsden2016stereotypes}. Most of the user scenarios written by UX experts described male tech workers, aiming to maximize their efficiency. 
As an illustrative example consider the user scenario by \username{aerooctopus} (UX expert, no writing assistant), who created the role of a investment banker Timothy Claude, who is in constant stress while their wife stays at home with their kids:

\begin{quote}
{\sffamily
    ``As part of his job as an \textcolor{blue}{investment banker}, he also desires the ability to stay focus for extended periods, he desires to have this feature present on the app. He also \textcolor{blue}{hates moving around with his umbrella}, he desires a feature that can tell the weather condition at a particular time. This will also assist him to know if to leave the house with his sunglasses everyday. \textcolor{blue}{Also, he was hoping if there is a way to include, remote monitoring of his home in the app, which will enable him see what is happening at home, especially when the kids are at home with his spouse.}''
}
\end{quote}

We acknowledge that the predominance of user scenarios focusing on male workers in technology may be influenced by the participant sample (21 out of 30 self-identified as men). However, we still observe significantly more variety in the user scenarios written by laypeople, despite a similar gender distribution.

\subsubsection*{\textbf{$RQ_{1}$: Are laypersons able to write reasonable user scenarios?}}
\label{sec:evaluation}

By reviewing the user scenarios, we found that laypeople can be empowered to write meaningful and cohesive user scenarios that also mainly follow the desired format, including extensive and relatable role introductions, routines, challenges, and goals. This is especially true for audience-orientation and structure (refer to Figure \ref{fig:boxplots}), where our data show a great difference regarding the use of a writing assistance. 

We discovered several essential characteristics of both strong and weak user scenarios, which were present in those created by experts as well as those written by laypeople. While the main issues of user scenarios by laypeople were primarily structure-related, such as too specific or too long scenarios, they overall performed better in the audience-orientation and effect categories. 

Strong user scenarios are relatable and easy to grasp, audience-oriented (they consider other stakeholders), clear and unambiguous. They share an authentic perspective based on a meaningful role, and clearly identified core challenges and suitable goals.

Poorly rated user scenarios show a lack of structure (they are cluttered, too long, or too broad), and are written without the audience in mind. They contain no role introduction or cannot comprehensibly express the role, which often implies that challenges or goals are not comprehensible. In many cases, they often did not define any challenges or goals at all, or---in contrast---were simply collections of needs and goals without specifying the necessary context.

\subsubsection*{\textbf{Influence of Writing Assistant}}
\label{sec:wa_influence}

When designing LLM-supported writing assistants, another key question is whether and to what extent the suggestions of the writing assistant had influenced the user scenarios at the time they were written. 
To evaluate the possible influence, we compared all suggestions from the writing assistant with the user scenarios to find text passages that were related to one or more suggestions. Once such a relation was identified, we investigated whether the passage was influenced by a suggestion. Due to the extensive logging, we were also able to identify these influences for intermediate user scenarios (i.e., work in progress text).

\begin{figure*}[t]
\includegraphics[width=\textwidth]{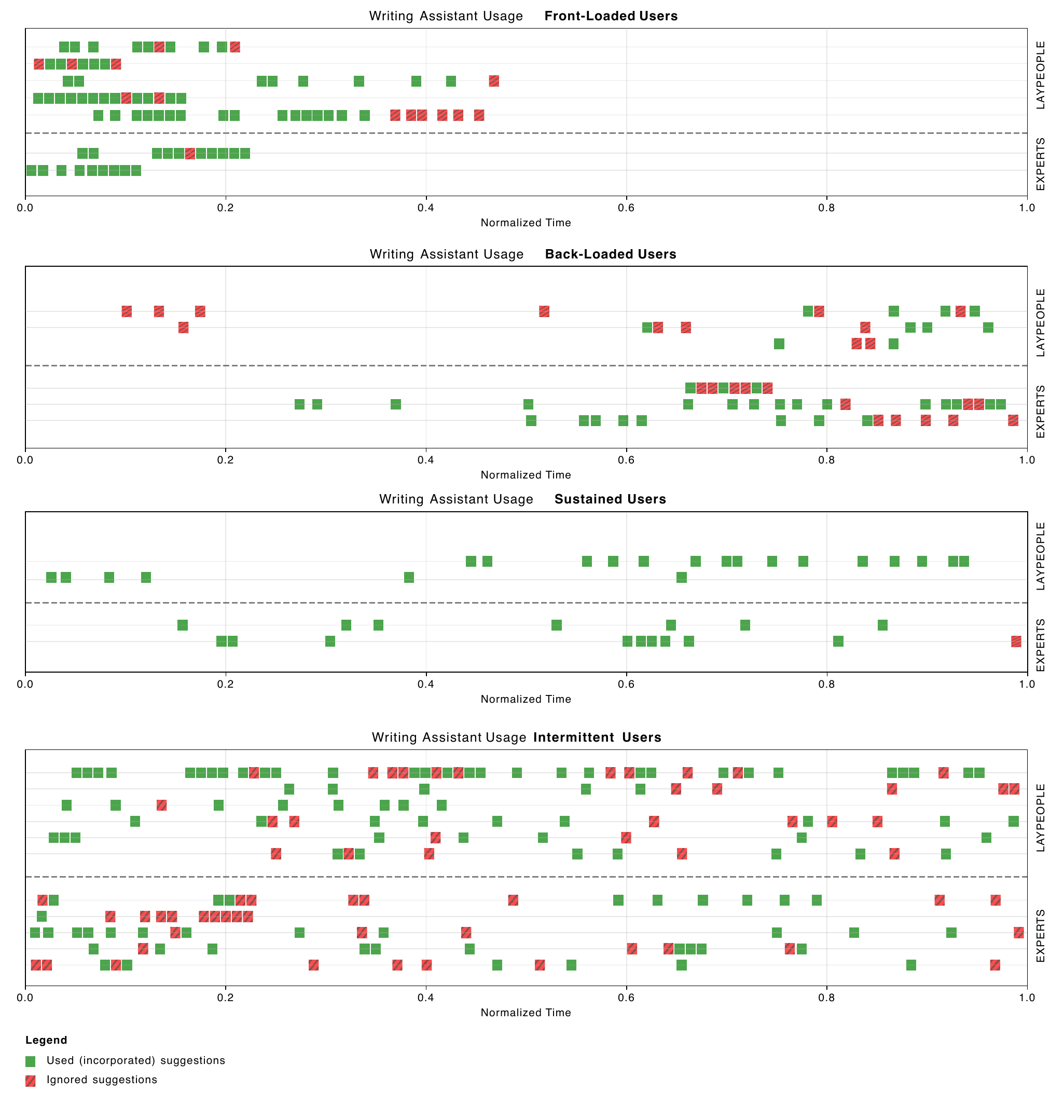}
\caption{Comparison of writing assistant usage between UX experts and laypeople showing when a participant used the suggestions or ignored or rejected them over time according to the procedure as described at the beginning of this section. All timestamps were normalized. Gaps in between either indicate idle time or writing without invoking the assistant. Note that timestamps were slightly modified to avoid overlaps for the purpose of a clean visualization.}
\Description{TODO}
 \label{fig:interaction_patterns}
\end{figure*}

After comparing all user scenarios and their corresponding logs, we ultimately found a total of 105 influences for 31 of 40 participants, as expressed by text passages that were added directly after a proposed suggestion. A visualization can be found in Figure \ref{fig:interaction_patterns}. From this inspection, we observed several common patterns of using the writing assistant. Some participants, such as \username{directionorillia} or \username{zaffreminidisc},  primarily invoked the writing assistant at the beginning of their writing process, while others had used it throughout the entire process (e.g., \username{silvermag}, \username{helenareview}). 

\begin{figure*}[t]
  \includegraphics[width=\textwidth]{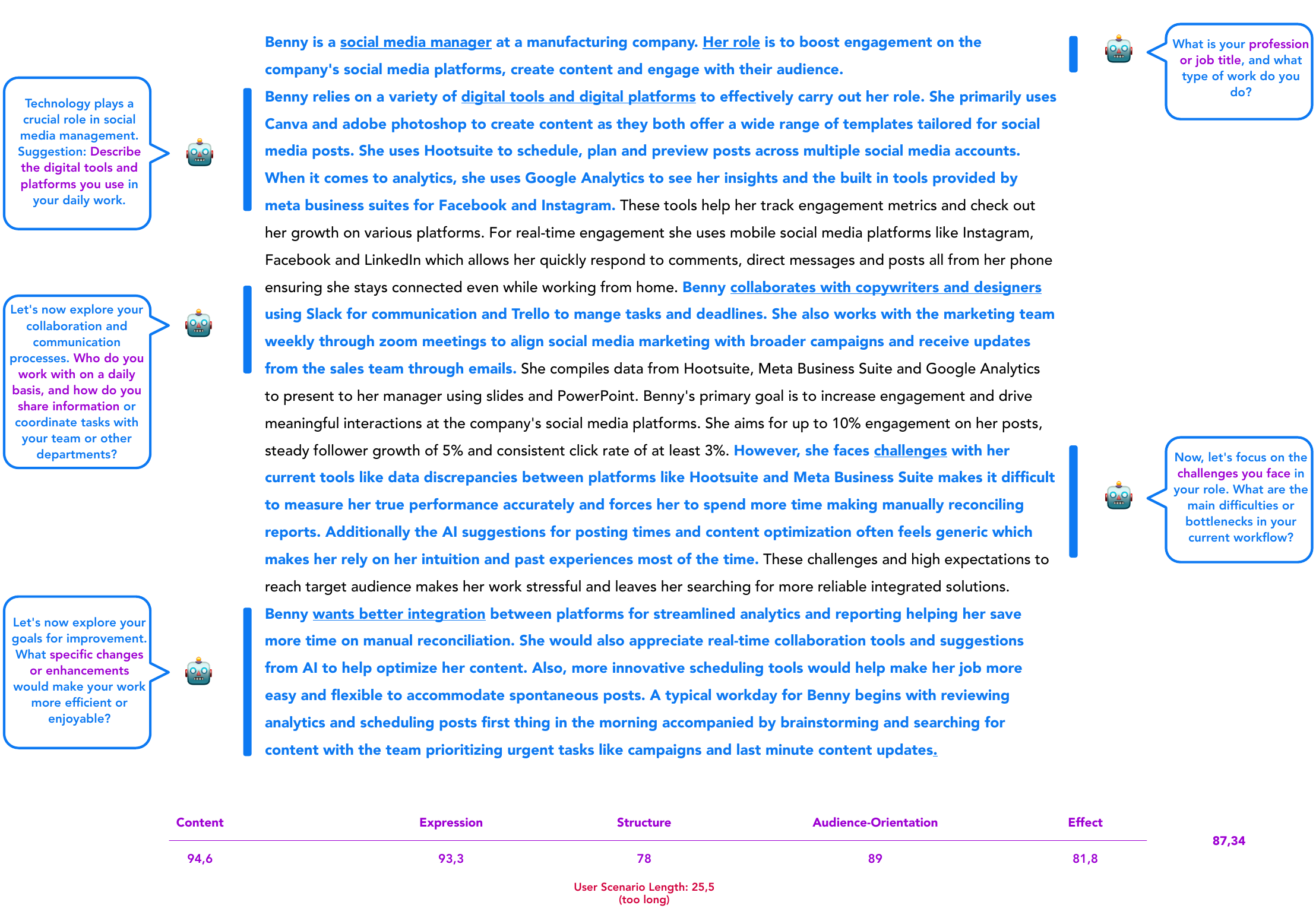}
  \caption{Visualization of the user scenario written by UX expert \username{shepherdornate} with the zero-shot writing assistant, including all incorporated suggestions. The blue highlighted segments were written after the participant had invoked the writing assistant. It is clearly recognizable that the suggestions had an influence on the user scenario.}
  \Description{TODO}
 \label{fig:suggestions1}
\end{figure*}

Figure \ref{fig:suggestions1} shows a visualization of the user scenario written by UX expert \username{shepherdornate} that shows multiple passages where the writing assistant had influenced the user scenario. This is also illustrated by the comments of this participant: ``It was extremely helpful as it helped me build on the story step by step'', pointing out that this participant used the suggestions as an anchor to guide them through the process and giving them the right ideas on what should follow next. 

The user scenario by this participant was rated high for most of the categories with only a single outlier: the length. Both raters agreed that the user scenario was too long. In fact, a lot of the smaller details and paragraphs could be paraphrased or reduced, which would be an important feature for an improved writing assistant that will also suggest to trim the text or cut certain paragraphs that are unnecessary (refer to Section \ref{sec:editing_deleting} for a brief discussion on this topic).


\paragraph{Chatbot-Like Behavior}
One particularly interesting way the writing assistant had influenced some participants can be described as a chatbot-like behavior. Some participants interacted with the LLM-supported writing assistant as if they were collaborating with a chatbot, although the writing assistant was explicitly instructed not to address users directly. This led to the point where the suggestions were treated as the beginning of a conversation between the writing assistant and the participant.
For instance, at one point, the writing assistant asked a participant if they could describe existing tools of their current workflow, and the participant responded directly to that question in the user scenario that there indeed was an app:

\begin{quote}
{\sffamily
\textbf{Assistant:} Great progress so far! Let's delve deeper into potential solutions. Could you describe any existing digital tools or systems your firm has tried to implement to address these challenges?

\textbf{Participant:} \textcolor{blue}{Yes there is an app} that every department can log on to message one another and send precise request with pictures and also alert if that task is completed
}
\end{quote}

In addition, some participants often responded with acknowledge words like ``Yes'' or ``okay'' to the suggestions before writing the corresponding section. Two participants (\username{mistakeglaucester}, layperson with only 1-2 years of experience with AI, and \username{mauvelouscouplekiss}, UX expert with also only 1-2 years of experience with AI) even wrote a final statement at the end of their user scenarios stating that they were done with their work, such as ``I think that is all'', after frequently using the writing assistant throughout the experiment.

\begin{quote}
{\sffamily
    \textbf{Participant:} \textcolor{blue}{No we dont have any particular problem solving tools} .I review most of the work been done by my subordinates. [..] . I responsible for software. We use google work place in most of our works. We usually collaborate. \textcolor{blue}{For now i think we are good.}
}
\end{quote}

Finding such acknowledgment phrases in the submitted user scenarios seems rather odd, as the user interface clearly indicated the setup: a text editor for writing the user scenario and the writing assistant as a distinct and decoupled module located at the bottom of the screen. However, these participants appear to have developed a 1:1 relationship with the writing assistant and its suggestions, resulting in direct responses within the text.

A particular example where a back-and-forth between suggesting and responding can be found in the following user scenario written by a layperson:

\begin{quote}
{\sffamily
    \textbf{Participant:} \textcolor{purple}{[Suggestion]} \textcolor{blue}{Okay} the goal of the firm is to get work done easy and firm ways to get more clients to work with. \textcolor{purple}{[Suggestion]} \textcolor{blue}{Okay} i handle review construction sites and checking work done of constructors to make sure their are going according to plan . \textcolor{purple}{[Suggestion]} \textcolor{blue}{Okay} when i arrive at work i go through the project files and check my meeting schedules for the day and know the times i have them , and start my day with reviewing the documents on my table . \textcolor{purple}{[Suggestion]} \textcolor{blue}{Okay} the firm logs on to the google workspace to work on their daily duties and community and have meetings using teams
}
\end{quote}

In this particular example, we can observe the participant directly responding to the suggestions by acknowledging first before they continue writing the actual part of the user scenario.

We also observed that some participants responded directly to the suggestions in the user scenario without embedding the suggestion with the answer in the text. This results in text passages that are difficult to understand because the original question or suggestion is ``lost''.

For example, taking this excerpt of the user scenario written by layperson \username{lancasteremergency} (zero-shot writing assistant):

\begin{quote}
{\sffamily
``Automated, non-biased, opt-in for the automated system, so it doesn't ping people who don't want it.

\textcolor{blue}{I'd say it should be twice a day, prioritized to members of my team. It should integrate with Slack, to automatically send to them.}

For the pairing solution, it should be weekly, prioritized to members outside of my team, and should integrate with JIRA to put them on the ticket as well, since they helped to pair.''
}
\end{quote}

It is nearly impossible to understand what the author is referring to in the highlighted paragraph, although the excerpt contains the surrounding context. 
It only becomes clear when we review the corresponding suggestion

\begin{quote}
{\sffamily
``Can you elaborate on how you envision this system working? Consider aspects like: 1. Frequency of reminders [..]''.
}
\end{quote}

\subsection{\textbf{Comparison of Zero-Shot and Few-Shot Writing Assistant}}
\label{sec:wa_zero_shot_few_shot}

To examine whether the presence of the writing assistant improved user scenario writing, and whether there are differences between the zero-shot and few-shot variants of the writing assistant, we conducted a robust two-way ANOVA with trimmed means based on the results of a preceding Levene's test for homogeneity of variance to verify equal variances across groups which slightly violated the assumption of homogeneous variances ($p = 0.071$), suggesting the use of a robust alternative.

\begin{figure}
  \includegraphics[width=\columnwidth]{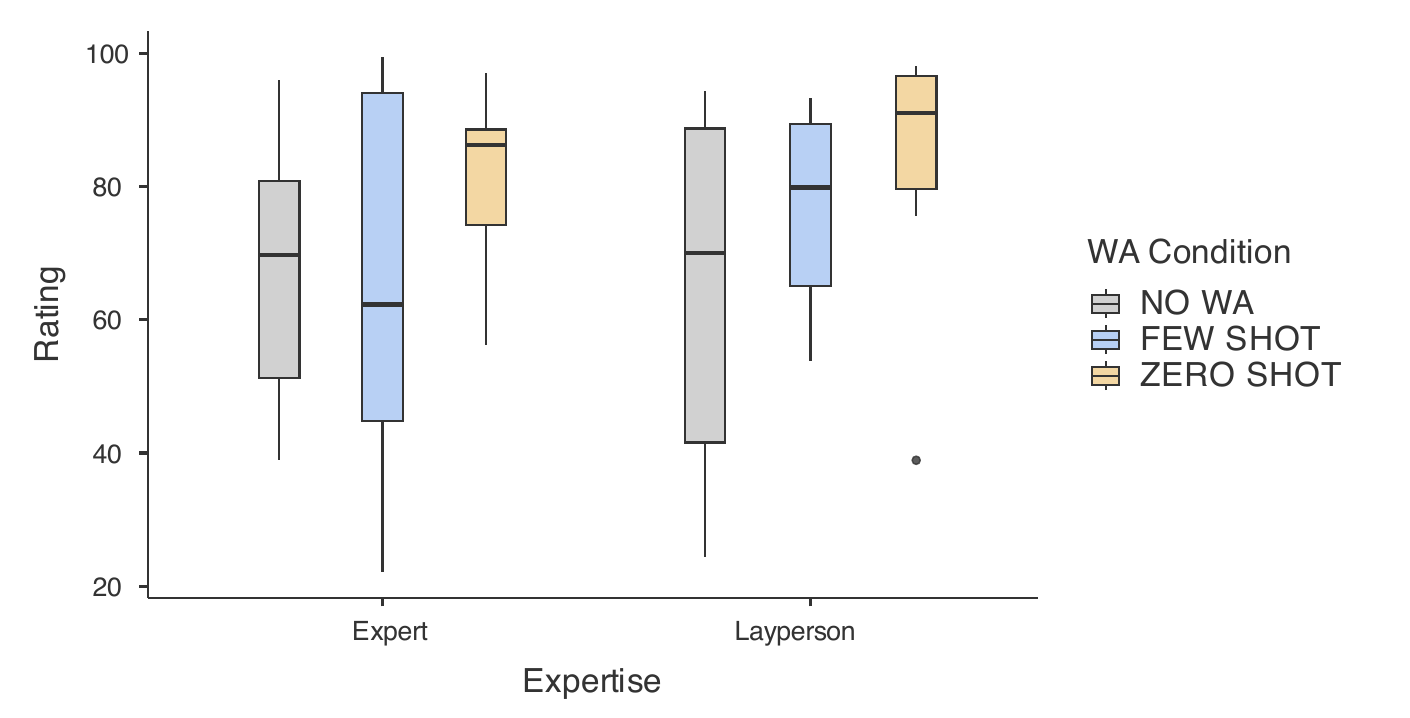}
  \caption{Boxplot comparison between UX experts and laypeople for their overall rating based on the writing assistant condition. Experts show less variability in ratings across writing assistant conditions, while the laypeople ratings clearly improved when using a writing assistant.}
  \Description{TODO}
 \label{fig:boxplots1}
\end{figure}

For this purpose, we used R's\footnote{https://www.r-project.org/} \texttt{bwtrim}\footnote{https://www.rdocumentation.org/packages/WRS2/versions/1.1-6/topics/bwtrim} function with rating as the dependent variable, and expertise (UX expert, layperson) and the writing assistant condition (none, zero-shot, and few-shot) as the independent variables.
We found a significant main effect for writing assistant condition ($p=0.0018$) indicating that the existence of a writing assistant had a significant impact on the user scenario quality regarding their rating\footnote{Only the zero-shot assistant showed a significant effect, but the lack of difference for the few-shot assistant may be due to sample size or participant variability rather than true ineffectiveness. This is discussed in the rest of this chapter.}. No significant main effect was observed for expertise ($p=0.5831$), nor was there a significant interaction between expertise and writing assistant condition ($p=0.3719$).

\begin{figure}
  \includegraphics[width=\columnwidth]{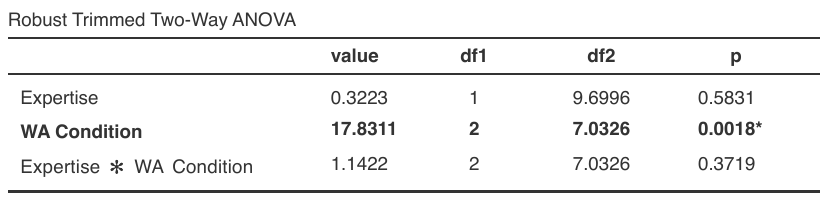}
  \caption{Results of a two-way ANOVA for expertise (UX expert, laypeople) and condition (no writing assistant, zero-shot, and few-shot writing assistant). We found a statistically significant difference between users with and without the writing assistant.}
  \Description{TODO}
 \label{fig:anova1}
\end{figure}

Post-hoc Tukey comparison revealed a significant difference between the zero-shot and none conditions ($p<0.05$), with higher ratings observed in the zero-shot condition compared to no assistant at all. However, no significant differences were found between the few-shot and zero-shot conditions or between the few-shot and none conditions ($p>0.05$).

\begin{figure}
  \includegraphics[width=\columnwidth]{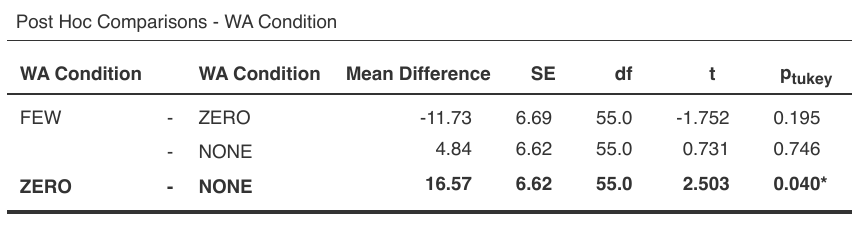}
  \caption{Results of a post-hoc comparison show a significant difference between users with the zero-shot writing assistant (ZERO) and users without any write assistant (none). We found no significant differences between zero-shot and few-shot (FEW) writing assistants, and between few-shot and none.}
  \Description{TODO}
 \label{fig:anova2}
\end{figure}

It should be noted that although there is a statistically significant difference with regard to the use of a writing assistant, this must be assessed as preliminary, especially in view to the fact that only the zero-shot variant is significant, and further investigation is necessary in subsequent studies. Given an effect size of $\eta^2=0.104$, we infer a small-to-moderate effect for the existence of a writing assistant, but expertise and the interaction between expertise and assistant type only show a minimal impact.

\subsubsection*{\textbf{$RQ_{2}$: Are There Differences Between Zero-Shot and Few-Shot Approaches?}}
\label{sec:r2}

Our findings indicate that the availability of an LLM-supported writing assistant demonstratively had an impact, with the zero-shot writing assistant making a significant difference.
The ANOVA showed a significant effect of the writing assistant condition, mainly due to the zero-shot assistant, which significantly outperformed the no-assistant conditions (A and baseline). The few-shot assistant's performance did not significantly differ from the zero-shot or no-assistant groups. This doesn't necessarily indicate that the few-shot assistant was less effective; rather, it may be due to participant variability or insufficient power to detect smaller effects. The strong performance of the zero-shot assistant might be influenced by specific participant sample characteristics, suggesting the need for further research with a larger or targeted sample to better understand the impact of the two assistant types.

Another reason why the differences between the few-shot and zero-shot suggestions are rather small and the zero-shot write assistant performing better than the few-shot assistant could also be the way we sampled the examples. Rather than primarily depending on existing user scenarios from online sources, a more refined write assistant should examine whether concrete and detailed descriptions of subsequent steps would improve the few-shot writing assistant, by gathering intermediate data from an actual setup that better demonstrates the rationale behind specific steps in the process. In addition, as depicted in Figure \ref{fig:boxplots1}, UX experts performed worse using the few-shot writing assistant than in the other conditions, which supports our assumption that constantly adding suggestions can have a negative impact on user scenarios.
Although both approaches, the zero-shot and the few-shot, were perceived as helpful by the participants, we found some indications that the few-shot suggestions were sometimes too specific and tailored to the data; however, also often more complex. Few-shot prompts are more involved in problem-solving, offering practical advice and prompting users to consider solutions. In contrast, zero-shot prompts aim to guide the users in a more abstract manner without necessarily steering toward particular details, such as concrete solutions or improvements. Refer to Table \ref{tab:wa_comp_zs_fs} for a brief overview of some zero-shot and few-shot examples. Both approaches had their benefits but also disadvantages. While the few-shot suggestions were much more detailed and aimed to dig deeper into certain aspects, they sometimes forced the participants to continue adding more and more minor details to their user scenarios, which sometimes resulted in more bloated and more complex scenarios.
An example few-shot suggestion taken from the study asks the participant to sketch out an ideal platform to enhance their workflow. While this forced the participant to develop an idea, possibly revisiting existing challenges, the approach also blurs some of the necessary abstraction. 

\begin{quote}
{\sffamily
    \textbf{Assistant}: ``Let's explore how technology could enhance your workflow. Can you envision an ideal digital platform that would integrate project management, communication, and design review processes?'' 
}
\end{quote}

On the other hand, the few-shot suggestions often asked for possible improvements for specific challenges the participants mentioned in their user scenario, which creates a stronger relation between the individual segments of the scenario and invites them to revisit some of these challenges to come up with possible improvements. In contrast, the zero-shot suggestions were rather straightforward and focused on the key aspects defined in the pre-prompts, such as the description of challenges or goals, without going into too much detail.



\subsection{Assessment of Generated Use Cases}
\label{sec:use_cases}
Overall, the generated use cases were rated high with an average rating of $86.05$. The LLM identified important aspects, such as actors, pre- and post-conditions, and the necessary steps, mostly correct and concise. 

\subsubsection*{\textbf{Extracting Actors}}
One difficulty encountered was identifying primary or secondary actors and their interconnections, often leading to overly specific actors. Notably, the role description significantly impacted the information extracted. Although technically correct, i.e. the LLMs successfully extracted the primary actor from the scenario, one should consider abstracting or separating the user scenario from the primary or secondary actors for a more generalizable perspective. How specifically the actors were named depended primarily on how the role was presented and whether a name for the persona was described there. While the LLM did a great job at determining the actors for certain user scenarios, there are still many edge cases or more complex cases that remain a challenge \cite{al2018use, wang2024llms}.
For example, an excerpt from a role description for a scientific researcher will return \texttt{Researcher working in quantum chemistry} as the primary actor for the generated use cases. 

\begin{quote}
{\sffamily
    ``My work is in \textcolor{blue}{scientific research}, specifically \textcolor{blue}{quantum chemistry} and molecular simulations''
}
\end{quote}

A very similar example is that of Sarah, which will be converted into \texttt{Primary actor: Sarah, the busy professional} for the generated use cases.

\begin{quote}
{\sffamily
    ``\textcolor{blue}{Sarah, a 32-year-old busy professional}, is looking to improve her overall fitness''
}
\end{quote}

When no specific users are introduced, the LLM often hallucinated and assumed the primary actor, as observed in the case of a user scenario by culturaldigital (layperson). They mentioned some technology to improve a system's security without stating any personnel, resulting in an assumption from the LLM that the primary actor would probably be a security expert. Although that assumption is not especially wrong, it was never stated in the original user scenario.

\begin{quote}
{\sffamily
    ``One last technology is adding some form of Blockchain to improve the \textcolor{blue}{security} and user individuality, while also pledging to secure and protect each users data''
}
\end{quote}

The LLM also generated nonhuman actors, such as \texttt{System process}, although it should be obvious from the user scenario that the entire described system will probably be developed by a team of software engineers covering that particular use case.

\begin{quote}
{\sffamily
    ``\textcolor{blue}{A back-end process} should be created to capture the name and e-mail collected from the end user''
}
\end{quote}

\subsubsection*{\textbf{Irrelevant Use Cases}}
Another challenge we found were superfluous or irrelevant use cases and meta-use cases. 

For instance, given the very clear suggestions on how to improve the daily work routine of a STEM teacher (\username{helenareviev}, layperson), a skilled UX expert could easily extract some goals for a new system, for instance, improving the way administrative tasks are handled.

\begin{quote}
{\sffamily
    ``\highlighted[JungleGreen]{black}{Routine} \textcolor{JungleGreen}{I start by opening Canvas, I got to the module that I am teaching and look at the content there, I download lessons and materials}
    [..]
    \highlighted[teal]{black}{Goals} \textcolor{teal}{My goals for improvement would be that I could focus more on the creative aspects for planning and development without having to focus so much on the administrative daily tasks that are required for each student. I'd like to streamline larger task into manageable steps}''
}
\end{quote}

However, one of the generated use cases described how to open a certain software for lesson preparation, which was introduced by the author when describing their daily routine. This step was only a descriptive element and was not meant to be part of the system design.

\begin{quote}
{\sffamily
    \textbf{Use Case 1: Opening Canvas for Lesson Preparation}\\
    This use case describes how an educator accesses the Canvas platform to begin preparing for their daily lessons, ensuring they have all the necessary content for their module.
}
\end{quote}

Although technically, the use case is valid as it reflects a real part of the user scenario, its value for requirements analysis is rather minimal, leading to unnecessarily bloated use cases. It appears that the LLM still lacks a deeper understanding of entities and relations between the individual segments of a user scenario. Current research suggested various ways of how to create such relations to extract the core aspects of use cases, such as assigning entities for all the mentioned goals \cite{gilson2020generating}. 







\subsubsection*{\textbf{Reception of Use Case Preview}}
Participants were optionally able to generate use cases from their user scenarios using the preview feature. This feature was intended not only as a writing aid, but also as a form of reflection, helping to clarify or refine the user scenario.

``The AI writing assistant was fairly useful in analyzing the general synopsis of a given text, and generating a set of use cases. I thought considering the descriptions of these scenarios were fairly short, the AI did a fairly good job with describing these use cases in more detail.'' (\username{finishbismarck}, layperson)

``I'll admit I didn't use the writing assistant that much, mostly because it was just as easy to write a scenario and check the Use Case Preview.'' (\username{tangerineghost}, UX expert)

\subsubsection*{\textbf{$RQ_{3}$: How useful are LLM-generated use case previews in user scenario writing?}} 
The idea behind previewing LLM-generated use cases was to provide the user with an additional representation to indicate how well a user scenario transformed into a more formal abstraction, and to help them parse and refine the scenario faster, spotting unclear or ambiguous elements better \cite{gallardo2007use}. The implementation did not rely on the writing assistant, meaning no implicit feedback mechanism was integrated to channel use cases back into the writing assistant. As illustrated in this section, this could significantly impact two aspects: a) enhancing relations among various actors, goals, etc., and b) to ask for clarifications in the user scenario when certain aspects seem unclear. Ultimately, the used LLM was found to generate reasonable use cases which could be optimized by a lot to be useful to a third party (e.g., a designer or software engineer); however, from the perspective of assistance in writing the user scenarios, their impact appears to be less relevant as an additional tool in the writing process itself.

\subsection{Reception of the Writing Assistant}
Overall, the writing assistant received positive feedback from most participants. We grouped and summarized most of the feedback in the following section.

\subsubsection*{\textbf{Bootstrapping the Process}}
A primary function of the writing assistant was to initiate the writing process. It offered suggestions, such as starting points or information on the profession and the daily work routine. This enabled users to begin writing even when they lacked initial ideas. A majority of participants used the writing assistant at the very beginning of their writing process.

We observed that many participants used the writing assistant to bootstrap their user scenarios, as laypersons were especially unfamiliar with the anatomy and structure of those user scenarios. Beginning to write the user scenario appeared to be a bigger issue for many participants, but ultimately all participants could write the user scenario after they had requested some suggestions. Some participants mentioned they liked how the writing assistant enabled them to start the process; they ``did not even know where to begin, or where to go'' and that the ``assistant helped a LOT'' (\username{mistakegloucester}, layperson), or that it even ``would save a ton of time when writing scenarios in my actual day job'' (\username{zaffreminidisc}, UX expert).

\subsubsection*{\textbf{Facilitating the Writing Process}} 
Many participants stated that the writing assistant helped them to facilitate the writing process. Writing a user scenario requires some logical thinking and organization as one puts together the different segments, such as extracting and writing down the problem definition. One participant (\username{ferrypsychology}, layperson) even compared the process to learning how to write an essay in school. They stated that the writing assistant ``made it an easy 1-2-3 process for writing a paper that would've taken me an hour to get to the same quality''.

Often mentioned aspects were that the writing assistant helped them to focus on important aspects by pointing out where they missed on, or how to continue writing (\username{cartwheelgainsboro}, UX expert), or by providing logical next steps (\username{nowcroquet}, layperson).

\subsubsection*{\textbf{Failing to Get 100\% Completeness}}
\label{sec:failing_completeness}
Some participants clearly noted they found it challenging to achieve 100\% completeness. It should be noted, that this was not a communicated goal or encouraged by our system.
As outlined in Section \ref{llm-based-writing-assistant}, an LLM was used to assess completeness  employing a non-deterministic method. At a certain point, the writing assistant would keep providing minor suggestions, never fully satisfied. Certain participants noted they couldn't reach 100\% completion due to character limits; one participant (\username{baghdadpayment}, layperson) mentioned feeling the completion grade was initially helpful, but nearing 95\%, AI suggestions seemed trivial and didn't seem to enhance the story significantly. They also hit the character limit before refining the story to 100\% completeness.
One participant (\username{tophatsunglow}, UX expert) repeatedly improved their user scenario using the writing assistant, but eventually hit the character limit. Initially, the scenario covered essential details like role, routine, problem, and workflow challenges. Despite reaching the limit, the participant kept using the assistant's suggestions until the scenario became overly complex. This case highlights the potential negative effects of the writing assistant when aiming for thorough completion.
A notable point from a participant was that the writing assistant never advised removing elements from the user scenario, only suggesting additions. This process seemed to trigger a cycle of excessive detailing, resulting in a flawed user scenario.




\section{Discussion}\label{sec:discussion}
Writing meaningful user scenarios is a non-trivial task. 
Researchers and developers could benefit greatly from the first-hand insights from actual users of a potential system, sharing their daily routines and real challenges and pain points that occur within their work day and when using their setup (tools, devices, etc.). A good user scenario needs to be balanced, i.e., contain enough details to learn about the user and to extract goals and user needs while also avoiding feature creep or being too complex and bloated. User scenarios are not meant to be an isolated artifact decoupled from the rest of the requirements analysis process and thus should be comprehensible by other stakeholders, such as UX designers. 
In this section, we now want to classify some of the results of our study.

\subsubsection*{Empowering Laypeople to Write User Scenarios}
One of the major chances for involving laypersons in the requirements analysis process is the uniqueness of the perspective these potential users can share by creating first-hand experiences and addressing real and concrete challenges. While experienced UX experts were able to create user scenarios that were audience-oriented, relatable, and memorable, despite they had to invent a persona and some story, our results also indicate that this heavily relies on the individual expertise and ability to empathize with users. In contrast, many other UX experts often created user scenarios that received low audience-orientation ratings. In this case laypersons really could make a notable difference.

By providing meaningful next steps while leaving concrete details to the participant, the submitted user scenarios follow the right structure and cover many insights that are relatable to the audience. However, it should be obvious that the ultimate result still greatly depends on the individual participant, as using the provided writing assistant did not enable every participant to write a good user scenario.

Our evaluation indicates that there are often subtle differences between the user scenarios, such as more detailed daily routine descriptions; however, in general, both groups, the UX experts and the laypersons, seem to have followed a similar structure. Notably, participants without the LLM-supported writing assistant often missed sections, such as describing challenges or daily routine, and were often comprised of a brief role introduction followed by multiple enumerations of desired goals.

There are some common issues with the low-rated user scenarios that could be summarized as a lack of meaningful personas and context. User scenarios lacking an introduction or insights into daily routines are hard to relate to, making it difficult to understand their specific challenges. Most of the weaker user scenarios had incoherent or missing roles, showed no challenges or pain points, and were often sole collections of wishes and requirements without any further context of why they would need those or what they aimed to improve.

\subsubsection*{Performance of LLMs}
One result that may seem a little counter-intuitive at first is the fact that the zero-shot variant performed better than the few-shot variant. Even if there were no significant differences between the two variants and the differences may also lie in the study design or the sample, they are similar to the results of \cite{kolthoff2024interlinking} where the zero-shot variant also outperformed few-shot.

Regarding the LLM-generated use cases based on the user scenarios, we encountered similar results to \cite{kc2024analysis}. One of the main challenges was to automatically extract the involved actors, understanding their relations to other occurring persons, and to differentiate between human and non-human actors, such as a system or a service. We showed some illustrative examples in Section \ref{sec:use_cases}. The quality of the generated use cases also depends on the user scenarios, although it appears to be more robust and tolerant to minor issues regarding the structure, for instance. However, certain aspects can improve the success of the generated use cases, such as using unambiguous language and the definition of clear actors and roles (compare \cite{bajaj2022muce}). Improved writing assistants should be instructed to consider these aspects for their suggestions.

\subsubsection*{Editing and Deleting Segments}
\label{sec:editing_deleting}
User scenarios written with the help of the LLM-supported writing assistant were intricate and detailed. Although much of the information offered valuable insights into challenges and daily routines, a significant drawback was that the writing assistant never advised shortening or removing certain elements while continuously suggesting additional details for existing segments. As highlighted in section \ref{sec:failing_completeness}, some participants tried to achieve a 100\% completeness level, which was, in fact, just an estimation by the LLM as well. Together with the technical nature of LLMs, this can easily create an infinite cycle of repetitions and suggestions to add more and more details --- unaware or inexperienced users could get stuck and lose the plot. 
An effective enhancement could involve incorporating specific editing guidelines in the prompts, guiding the LLM to propose modifications or even deletions of particular words or entire text segments \cite{chakrabarty2024can}.



\subsubsection*{Overusing Could ``Prevent'' Individualism}
The writing assistant was designed to help users create their own user scenarios, with the idea of enabling them instead of dictating. Therefore, suggestions provided were intended as prompts for reflection, without specific examples or pre-made text blocks for direct use. 
However, we observed that some participants showed that they had fully started to rely on the suggestions up to the point where they sometimes would invoke the writing assistant constantly before continuing the user scenario. One drawback of this observed behavior is that it may lead to more cluttered and “robotic” user scenarios, prevent individualism, and influence the realism of the user scenarios. This phenomenon of overreliance has been extensively discussed in prior research \cite{vasconcelos2023explanations, kim2024m}. 
A possible improvement for these problems could be to ensure more reasonable and reliable LLM outputs, for instance, by providing more in-depth explanations. In addition, using conversation memory or a state that stores the previous suggestions and locks them until a certain condition is fulfilled could help with repetitions. Another change could be slightly modifying the prompt to clarify that the previous suggestion should be seen as the foundation of the next suggestion rather than tricking the LLM into suggesting something different. 


\section{Limitations}\label{limitations}

\subsubsection*{\textbf{Submission quality and AI-generated content}} 
Although the overall quality of online studies is rated as positive \cite{douglas2023data}, we would like to address some limitations that occurred within this study.
We had to reject and exclude some submissions due to low-effort responses or the use of external AI (such as ChatGPT). As we were logging nearly every interaction with our system, we were able to identify suspicious-looking user scenarios rather easily, for instance, because of the way they were entered or copy-pasted into the editor or typical LLM generation artifacts or formatting. Although the majority of instances were quite clear, in certain situations, we had to dismiss and remove submissions as a precaution, even without definitive proof of them being AI-generated, to ensure a high-quality standard. However, when applying the right screeners, the overall quality was solid. 

\subsubsection*{\textbf{Different Initial Positions}}\label{different-initial-positions}
While laypersons were tasked with writing a user scenario from their personal point of view, UX experts were required to write from the perspective of someone else, such as a colleague or friend. This approach was chosen to collect diverse submissions, allowing an unbiased assessment and examination of the user scenarios and resulting use cases.

\subsubsection*{\textbf{Reliability of Ratings}}
\label{reliability-of-ratings}
To rate the user scenarios submitted, we assigned two researchers from our lab who were experienced in requirements analysis. Although the inter-rater reliability was sufficient, more reliable results could be achieved by employing several independent raters from the field, such as other UX experts. 

\section{Possible Applications and Future Work}
We believe that the individual and authentic perspective of laypeople and possible users of a system could be a crucial key to better software design and truly enrich the human-centered cycle. Genuine user scenarios contain lots of insightful data and perspectives---often between the lines---and should not be entirely replaced with artificial intelligence or outside experts. This could allow for closer collaboration between UX designers and laypeople. 
Another possible application we can imagine is a more tailored user adaptation in applications that will allow for applications that fit the individual user and their needs, challenges, and goals. Such user scenarios can ultimately be transformed from the written narrative to specific use cases. 
Future work should evaluate more robust approaches to supported user scenario writing by proactively suggesting further refinements, and by capturing common issues and providing meaningful suggestions to fix those.

\section{Conclusion}\label{discussion-and-outlook}
This paper presents a between-subjects comparison between UX experts and laypersons regarding the writing of user scenarios with or without the support of a LLM-based writing assistant. We specifically examined whether laypersons can write meaningful user scenarios. We highlighted differences in how these participants approached the process and provided design rationales for writing assistants. Our study indicates that both UX experts and non-experts encountered challenges and succeeded in this task to a similar extent. However, our findings also suggest that laypersons can be empowered to write good user scenarios. In particular, fourteen out of the 25 best-rated user scenarios were crafted by laypersons, and twelve of those had LLM-supported writing assistance. The best-rated user scenario was authored by a UX expert using a writing assistant (\username{ebonyballoon}), yet the first-hand scenario by layperson \username{canmorecontract} was rated with similar quality. While we could not quantitatively prove that the writing assistant increased the overall quality of the user scenarios for every participant, we showed that the existence of a writing assistant had a significant influence on the user scenarios and that such assistants can in turn help to improve certain aspects of a user scenario and probably enabled several participants to write a meaningful user scenario. Our study design does not allow us to determine how user scenarios written with LLM-supported assistance could have come out without the writing assistant; however, receiving mostly positive comments from laypersons, noting that the assistant facilitated starting the process, staying focused, and selecting suitable topics and words, among other things, can be seen as an indication that the writing assistant was helpful to write the user scenarios. On the other hand, LLM suggestions can create artifacts, such as confirmation words, and increase the length or complexity when providing indications or assumptions on the completeness, and sometimes these suggestions took away some of the individual freedom because participants became too accustomed to the suggestions.


\bibliographystyle{ACM-Reference-Format}
\bibliography{main}


\appendix

\clearpage
\onecolumn
\section{User Study Material}

\subsection{Example User Scenario Taken in Few-Shot Examples}
\label{sec:appendix_user_scenario}
\vspace{0.5em}
\begin{mdframed}[backgroundcolor=gray!10, roundcorner=5pt]
\small
\vspace{0.5em}
\noindent\textbf{User Scenario} \\
As a professional photographer, I want a comprehensive system that will revolutionize my entire workflow, from client management to photo editing and delivery. The system should include a robust CRM, advanced AI-powered editing tools, and seamless integration with all major social media platforms. It must be able to handle RAW files from any camera brand, automatically organize my photos, and suggest the best edits based on current trends. [..] This all-in-one solution will save me countless hours and streamline my entire photography business.
\end{mdframed}
\vspace{0.5em}

\subsection{Rating Items}

\subsubsection{Rating items for user scenarios}
\label{tab:rating_items_us}

\begin{mdframed}[backgroundcolor=gray!10, roundcorner=5pt]
\small
\vspace{0.5em}
\textbf{Content}\\ Does the scenario base on real data?\\ Does the scenario reflect the feeling of the original events and the way in which the participants themselves might express it?\\ Does the scenario describe the way things happened thoroughly?\\ Does the scenario identify a point-of-pain/a market gap/a new approach/a trend that is relevant for the UX issue?\\ Are the characters and situations representative?\\ Is the content of the scenario specific and tangible?\\ \textbf{Expression}\\ Does the scenario have a clear focus/a clear point?\\ Does the scenario use just enough details to help the audience recognize its focus and authenticity, and no more?\\ Is the message of the scenario direct?\\ Does the scenario provide enough contextual detail to help the audience relate to it?\\ Does the scenario use the characters’ language?\\ Does the scenario use active descriptions?\\ \textbf{Structure}\\ Does the scenario address all of the facts, even inconvenient details?\\ Do the facts and explanations delivered in the scenario make sense?\\ Is the explanation/solution delivered in the scenario convincing?\\ Do the presented facts fit well in the scenario?\\ Is the scenario as long as it needs to be, but no longer?\\ \textbf{Audience-Orientation}\\ Is the audience able to complete the “journey” in their minds?\\ Is the scenario open-ended, so that it can be interpreted by the audience?\\ Is it told from a perspective appropriate to the audience?\\ Does the audience know enough about the context of the scenario?\\ \textbf{Effect}\\ Is the scenario memorable?\\ Can the audience identify to/empathize with the characters of the scenario?\\ Does the scenario inspire new ideas?
\end{mdframed}

\clearpage
\subsubsection{Rating items for use cases}
\label{tab:rating_items_uc}

\begin{mdframed}[backgroundcolor=gray!10, roundcorner=5pt]
\small
\vspace{0.5em}
Plausibility – the realism of the use case\\ Readability – the flow of the use case\\ Consistent structure – consistent terminology and use of present simple tense\\ Alternative flow – consideration of variations\\ Actors – the correct actors were identified. Correctness was determined relative to the informal requirements specification.\\ Use cases – the correct use cases were identified. Correctness was determined relative to the informal requirements specification as above.\\ Content – the description of each use case contained the information required by all the sets of guidelines: actor, assumptions that \\ must be valid before the use case starts, flow of events, variations and post-conditions.\\ Level of detail – the descriptions of each event were at an appropriate level of detail. \\ There should be no unnecessary details about user interface or internal design. \\ Each event should be atomic, that is, sentences with more than two clauses should be avoided.\\ Realism – the flow of events was realistic, that is, the events follow a logical and complete sequence, \\ and it is clearly stated where variations can occur.\\ Consistency – the use of terminology was consistent.
\end{mdframed}

\subsection{User Scenarios}

{\small
\begin{longtable}{|l|l|p{.8\textwidth}|}
\caption{Overview of all user scenarios} \label{tab:us_overview_experts} \\
\hline
\textbf{ID} & \textbf{Cond.} & \textbf{Title} \\ \hline
\endfirsthead
\hline
\textbf{ID} & \textbf{Cond.} & \textbf{Title} \\ \hline
\endhead

    koalabeige & B & Cross-platform user testing and bug reporting workflow for web application \\ \hline
    mauvelouscouplekiss & B & IT Training and Support for Older Employees to Enhance Workplace Technology Adoption \\ \hline
    cordovantriumph & Baseline & Marketing Manager Creates and Manages Holiday Promotional Campaign Using Company Platform \\ \hline
    beerscrimson & D & Professional Seeks Home Fitness Equipment: Easy Online Shopping for Treadmill and Dumbbells \\ \hline
    denimprojector & Baseline & Welcome popup with newsletter subscription for first-time visitors \\ \hline
    zaffreminidisc & D & Scientist seeks AI tool for experiment optimization, data analysis, and cross-functional collaboration \\ \hline
    tophatsunglow & D & Software Engineer aims to optimize workflow with tools and personal development goals \\ \hline
    cartwheelgainsboro & B & Interactive Mirror and Furniture E-commerce Website with Room Visualization Tool \\ \hline
    burritotan & B & Technology Engineer Optimizes Machine Control Systems for Company-Wide Efficiency \\ \hline
    tangerineghost & Baseline & Bakery Website Redesign: Enhancing Online Ordering and Inventory Management \\ \hline
    orangepager & B & Software Engineer Streamlines Debugging with All-in-One Platform \\ \hline
    aerooctopus & baseline & Investment Banker's All-in-One Smart Home and Health App \\ \hline
    silvermag & D & Web Application for Government Program Registration with Enhanced Security and Accessibility \\ \hline
    fountainlinen & B & Foodbay founder gathers diverse user data via online survey to shape food delivery platform \\ \hline
    ebonyballoon & D & College Student Finds Safe, Affordable Off-Campus Apartment Using User-Friendly Mobile App \\ \hline
    brainycoal & Baseline & Software Engineer Manages Client Requests and Development Challenges During Website Project \\ \hline
    zippyfrankincense & D & Photographer seeks improved portfolio platform with enhanced organization features \\ \hline
    darkyule & D & Web Designer Creates Bakery Website: From Concept to Client Approval \\ \hline
    presentsbright & D & 3D modeling of scary shark-car hybrid for horror movie set with specific tooth design \\ \hline
    drummervolcanic & Baseline & Data Analyst Creates Weekend Sales Report with Visualization Tool for E-commerce Company \\ \hline
    gingerbreadgaudy & B & IT Pro Seeks Unified Ecosystem: Streamlined Tools for Enhanced Productivity and Simplified Workflows \\ \hline
    wenceslaushandsome & D & Marketing Specialist Streamlines Email Campaigns with All-in-One Platform \\ \hline
    popcornglistening & Baseline & Marketing Manager Streamlines Campaigns with All-in-One Platform for Analytics, and Team Collaboration \\ \hline
    frostysullen & D & Web Designer Creates Responsive Nail Salon Website with Custom Color Palette and Multi-Device Preview \\ \hline
    shepherdornate & B & Social Media Manager Seeks Integrated Tools for Streamlined Analytics and Enhanced Content Optimization \\ \hline
    playerspectacle & B & Stay-at-Home Dad's Automated Expense Tracker for Family Budget and Subscription Management \\ \hline
    gnatmat & Baseline & Front-end engineer designs pharmaceutical e-commerce web app with CI/CD and cloud deployment \\ \hline
    liftertobacco & B & Streamlined Workflow Platform for Freelance Graphic Designers: Integrating Task Management \\ \hline
    termsalesman & Baseline & Project Manager Creates and Refines "Prospects" Design for Client Presentation \\ \hline
    explodeavoid & Baseline & Web Designer Creates Senior Travel Blogger Website with Branding and Responsive Design \\ \hline
procedurehawaii         & A             & Nurse Manager seeks efficient system for staff management and patient care coordination                              \\ \hline
summersideawareness     & A             & Automated tool to compare, analyze, and optimize data analysis approaches across different platforms                 \\ \hline
swanseareference        & C             & Student Intern Designs Modern Payroll System for College's Hourly Workers                                            \\ \hline
canmorecontract         & A             & Restaurant Management System for Streamlined Operations and Enhanced Customer Experience                             \\ \hline
lancasteremergency      & C             & Software Engineer's Productivity Boost: Automated PR Reviews and Weekly Pairing System                               \\ \hline
mumbaifoundation        & C             & AI-Powered Platform for Integrated Clinical Practice and Medical Education                                           \\ \hline
extentdelhi             & A             & Intelligent CRM with AI-powered personalized sales follow-up and customer interaction analysis                       \\ \hline
beijingindividual       & E             & Product Designer Seeks All-in-One Platform for Streamlined Workflow and Enhanced Collaboration                       \\ \hline
finishbismarck          & A             & Systems Administrator integrates Google Maps data for enhanced weather impact visualization                          \\ \hline
helenareview            & C             & Remote STEM Teacher Seeks Streamlined Workflow for Grading, Planning, and Communication                              \\ \hline
mistakegloucester       & E             & Freelance Transcriptionist Seeks to Expand Business and Improve Audio Quality for Efficient, Accurate Transcriptions \\ \hline
osakaguest              & C             & Construction Project Manager's Workflow Optimization for Site Inspections and Team Communication                     \\ \hline
hoveinflation           & E             & Task Management System for Real-Time Team Communication and Progress Tracking                                        \\ \hline
kentvillebreath         & E             & Plant Auction and Live Selling App for Mobile and Web with Secure Payments and Buyer Protection                      \\ \hline
waterlootonight         & A             & Marketer Creates Compliant Logo for Grant Initiative Using Adobe Suite                                               \\ \hline
baghdadpayment          & C             & Automated CIF File Standardization and Conversion Tool for Streamlined Quantum Chemistry Workflows                   \\ \hline
directionorillia        & E             & Teacher's Daily Routine and Challenges: Balancing Technology, Collaboration, and Classroom Management                \\ \hline
pumpkinspritsail        & C             & Software QA Manager Seeks to Optimize Testing Process and Improve Cross-Team Collaboration                           \\ \hline
culturaldigital         & E             & Open-Source AI-Assisted System Architecture Generator with Community Collaboration and Blockchain Security           \\ \hline
attendcooing            & A             & Auto Adjuster's Comprehensive Case Management and Guidance System                                                    \\ \hline
enjointittering         & C             & Salesforce Admin Seeks Integrated Ecosystem for Enhanced CRM Management and Collaboration                            \\ \hline
lesserafrican           & E             & Computer Analyst Seeks Efficient Software for Automated Issue Detection and Streamlined Troubleshooting Process      \\ \hline
autopsyforsaken         & A             & Social Media Manager Executes \#MindMattersNow Campaign and Analyzes Performance for Mental Health Awareness         \\ \hline
nowcroquet              & C             & College essay tutor needs improved software for efficient and reliable essay reviews                                 \\ \hline
evenpester              & E             & Ethical Tie-Dye Business Owner's Daily Workflow and Order Fulfillment Process                                        \\ \hline
reviewbellow            & A             & AI-powered Video Creation App with Automated Functions and Cloud Integration                                         \\ \hline
mailboxvillage          & A             & AI-powered Real Estate Transaction Management System for Admins                                                      \\ \hline
japanreading            & C             & AI-powered Medical Dictation Tool for Efficient Patient Documentation and Improved Workflow                          \\ \hline
airplanemembership      & E             & Production Lead Seeks Improved Communication System for Efficient Event Setup and Coordination                       \\ \hline
ferrypsychology         & E             & Streamline Employee Onboarding: Automate Account Creation and Access Provisioning for Efficient IT Operations        \\ \hline
amphoradepth            & A             & Efficient, Bug-Free Software for Remote Work Productivity with Advanced Features and User-Centric Design             \\ \hline
\hline
\end{longtable}
}

\clearpage
\section{Prompts}
\label{sec:system_intructions_wa}

\subsubsection{System Prompt for the writing assistant}
\label{tab:wa_system_prompt}

\begin{mdframed}[backgroundcolor=gray!10, roundcorner=5pt]
\small
\vspace{0.5em}
You are an AI assistant tasked with guiding users to write a comprehensive user scenario about their work. \\ This user scenario should cover how they work, which technologies they use, and with whom they collaborate. \\ Your goal is to help the user create a detailed and insightful narrative that can later be converted into distinct use cases by another AI system.\\ \\  The users receive the following instructions:\\  \# Users - Task Instructions\\         \textgreater As part of digitization, new digital processes and applications are to be developed for your job/field of activity, \\         \textgreater or these are to be expanded using new methods. \\         \textgreater Imagine that you are part of this development process. \\         \textgreater To ensure that the applications can cover your work well, \\         \textgreater you should help to define as many requirements and functions as possible. \\         \textgreater To do this, it is necessary for you to describe your view of your activities \\         \textgreater and which functions would be important to you.\\ \\ You will guide the users to achieve the best results for their task.\\ \# Guidelines\\         1. Be specific daily tasks and processes without being to detailed about every little aspect\\         2. Include information about tools, software, and technologies used on a high-level perspective\\         3. Describe collaborations and interactions with team members or other departments\\         4. Highlight any pain points or challenges in the current workflow\\         5. Mention any goals or desired improvements in the work process\\         6. The user can only reply via buttons in the frontend. They have no option to response with free text.\\  \# Tone and Format instructions\\         - Highlight important keywords in a bold face using the respective HTML tags, e.g., \textless{}b\textgreater{}daily routine\textless{}/b\textgreater{}.\\         - Always address the user directly and never speak of them in the 3rd person. You are here to directly help the user.\\         - Do not repeat the prompt everytime you are asked. It is also not necessary to say Thanks or something similar everytime. \\         Stay focused on your goal. \\         - It is important that your responses are brief and below 300 characters long. \\         - Do not suggest all steps at once, but rather sequentially based on the content that will be provided in the messages.\\ \\ \# Further Instructions\\         - The user has no option to chat with you! You only make suggestions that the user may incorporate \\         into their user scenario and their writing process!\\         - As you guide the user through this process, remember that their scenario will be used to create distinct use cases. \\         Encourage them to think about different scenarios or situations that occur in their work, as these \\         can be valuable for identifying separate use cases.\\         - Aim for a comprehensive narrative that covers all the aspects mentioned above. \\         Feel free to ask for clarification or more details if needed to create a rich and informative user scenario.\\         - I will provide more and more of the already written user scenario and you will guide the user \\         step-by-step and show the next logical step.
\end{mdframed}

\clearpage
\subsubsection{Prompt ``Help with Story''}
\label{tab:wa_help_story}

\small{
\begin{mdframed}[backgroundcolor=gray!10, roundcorner=5pt]
\vspace{0.5em}
Analyze the provided user scenario. 
\noindent This is the user scenario that the user has already written. 
If nothing is written in those tags or if the tags are missing, help the user to start writing their scenario.

\noindent Now, let's guide the user through the thought process of creating their user scenario. Ask the user to consider the following aspects of their work:

\noindent1. If unclear, let the user provide a basic idea of what they work, what their profession is called, to start\\
\noindent2. Daily routine: What does a typical workday look like?\\
\noindent3. Main responsibilities: What are the primary tasks and projects they work on?\\
\noindent4. Tools and technologies: Which software, hardware, or platforms do they use regularly?\\
\noindent5. Collaboration: Who do they work with, and how do they communicate or share information?\\
\noindent6. Challenges: What are the main difficulties or bottlenecks in their current workflow?\\
\noindent7. Goals: What improvements or changes would make their work more efficient or enjoyable?

\noindent Suggest one of these steps at a time to the user.

\noindent Do not include any explanation or additional comments that are not part of your task.

\noindent Encourage the user to provide specific examples and anecdotes that illustrate their experiences. Remind them that the more detailed and concrete their scenario is, the better it will be for generating accurate use cases.

\noindent\#\# The User Scenario
\[...\]
\end{mdframed}
}

\subsubsection{Prompt ``Suggest next step''}
\label{tab:wa_next_step}

\small{
\begin{mdframed}[backgroundcolor=gray!10, roundcorner=5pt]
\vspace{0.5em}
Analyze the provided user scenario. This is the user scenario that the user has already written. 

\noindent Analyze the provided user scenario and suggest this as the next step:\\

\noindent - Refine or rephrase unclear sentences or expressions,\\
\noindent - Remind the user to think about their use or disuse of technology in their current work setting and what they like and dislike,\\
\noindent - Analyze the provided user scenario based and provide suggestions to improve it,\\
\noindent - Suggest to test the outcome of the user scenario by using the 'Preview' button in the editor,\\
\noindent - Remember to save regularly! You can do this using the corresponding button at the top right or by pressing the key combination CTRL+S (or CMD+S).,\\
\noindent - Suggest to click the 'Complete' button in the top right corner to check for the progress and completeness of the user scenario. This will also show some indicators and directions on how to complete your user scenario by hovering over the buttom after clicking it.

\end{mdframed}
}

\subsubsection{System prompt for the use case preview from the user scenarios}
\label{tab:uc_system_prompt}

\small{
\begin{mdframed}[backgroundcolor=gray!10, roundcorner=5pt]
\small
\vspace{0.5em}
You are tasked with converting a user scenario into use cases for an interconnected novel digital system. \\ \\     The user scenario describes a daily work routine that has been transformed into a narrative format. \\     Your job is to extract the key activities and interactions from this scenario and convert them into distinct, \\     interconnected use cases that will illustrate a novel interactive system of various applications, and interaction patterns. \\     The use case is described concretely and step by step from the developer's perspective.\\ \\     To complete this task, follow these steps:\\     1. Carefully read and analyze the user scenario. Identify the main actors (users or roles) involved in the scenario.\\     2. Break down the scenario into discrete activities or interactions that could form individual use cases. \\     Look for actions that involve user interaction with the system or significant system processes.\\     3. For each identified activity or interaction, create a use case containing the following information:\\         Title: Provide a clear, concise title for the use case\\         Actor: Identify the primary actor (user or role) involved\\         Description: Briefly describe the purpose of the use case\\         Preconditions: List any necessary conditions that must be true before the use case can be executed\\     4. Ensure that each use case is self-contained and focused on a single, specific interaction or process.\\     5. Review your use cases to ensure they accurately represent the activities described in the original user scenario \\         and that they form a cohesive system when considered together.\\ \\     Remember to focus on creating clear, concise, and interconnected use cases that accurately represent the \\     novel digital system described in the user scenario.\\     \\     \# EXAMPLES\\  Use case for meal delivery application\\             As a food lover, I want to place food orders directly through an app, so that I can conveniently \\             choose to pay now or upon delivery,\\             and ensure that my order is sent to the restaurant for preparation and delivery.\\              \\                        \textless{}h2\textgreater{}Use Case 1: Ordering a Meal Online\textless{}/h2\textgreater\\             \textless{}p\textgreater{}Individuals can use an app to place food orders directly to restaurants. \\             When the user places an order, they are prompted to pay through the app or pay when the food arrives. \\             Once that is confirmed, the restaurant will receive a request through their system. \\             The food will then be prepared, packaged, and delivered to the individual.\\             In this case, the app must be able to receive orders, process payments, and \\             communicate with the restaurant electronically.\textless{}/p\textgreater\\             \textless{}p\textgreater{}\textless{}b\textgreater{}System\textless{}/b\textgreater{}: Food delivery application\textless{}/p\textgreater\\             \textless{}p\textgreater{}\textless{}b\textgreater{}Primary actor\textless{}/b\textgreater{}: Customer ordering a meal\textless{}/p\textgreater \\ \\ \[...\]
\end{mdframed}
}

\section{Writing Assistant}

\begin{table}[H]
\caption{Some exemplary comparisons between the zero-shot and few-shot suggestions given by the writing assistant for several occasions}
\label{tab:wa_comp_zs_fs}
\resizebox{\textwidth}{!}{%
\begin{tabular}{|l|l|}
\hline
\textbf{Challenges Zero-Shot} & \textbf{Challenges Few-Shot} \\ \hline
\begin{tabular}[c]{@{}l@{}}Consider adding: Current challenges with the existing system. \\ Key features you'd like in the new system. \\ Benefits this system would bring to students and staff \\ Try expanding your scenario with these elements.\end{tabular} & \begin{tabular}[c]{@{}l@{}}Based on the information provided, let's explore the challenges \\ you face in your current workflow. \\ Can you describe any difficulties or bottlenecks you encounter in your transcription work? \\ This could include issues with time management, software limitations, \\ client communication, or managing temporary employees. \\ What aspects of your work do you find most challenging or time-consuming?\end{tabular} \\ \hline
\textbf{Role Zero-Shot} & \textbf{Role Few-Shot} \\ \hline
\begin{tabular}[c]{@{}l@{}}Your user scenario is developing well! \\ Let's refine some parts to make it clearer and more precise: \\ Refine descriptions: Consider rephrasing some sentences for clarity. \\ For example, "Reggie Field is a freelancer on Upwork" \\ could be more specific about his role.\end{tabular} & \begin{tabular}[c]{@{}l@{}}Here's a suggestion for the first step: Begin your user scenario by \\ introducing your role and main responsibilities. \\ Describe your typical work environment and \\ the primary goals of your position. \\ This will set the stage for a more detailed account of \\ your daily activities and challenges.\end{tabular} \\ \hline
\textbf{Goals Zero-Shot} & \textbf{Goals Few-Shot} \\ \hline
\begin{tabular}[c]{@{}l@{}}Thank you for sharing those challenges. \\ Now, let's focus on your goals and aspirations. \\ What improvements or changes would make your work as a technology engineer \\ more efficient or enjoyable? \\ Consider areas like process automation, skill development, \\ or team collaboration enhancements.\end{tabular} & \begin{tabular}[c]{@{}l@{}}Great job detailing your workflow and collaboration. \\ Now, let's explore the goals you have for improving your work process. \\ Consider: 1. What changes would make your experiments more efficient? \\ 2. How could data analysis and presentation be streamlined? \\ 3. Are there ways to enhance collaboration with your global team? \\ 4. What new technologies or tools could benefit your research? \\ Please share some specific improvements or changes that would make your work \\ as a scientist more effective or enjoyable.\end{tabular} \\ \hline
\textbf{Collaboration Zero-Shot} & \textbf{Collaboration Few-Shot} \\ \hline
\begin{tabular}[c]{@{}l@{}}Great! You've provided a detailed user scenario about Alex, a software engineer. \\ Let's explore the collaboration aspect of Alex's work to enrich the scenario further. \\ Who does Alex collaborate with in his daily work, and how do they \\ communicate or share information? \\ Consider mentioning specific team members, departments, or external partners.\end{tabular} & \begin{tabular}[c]{@{}l@{}}Add more details about collaboration. \\ You've mentioned coordinating with the band, venue staff, and production team. \\ Can you elaborate on how you communicate and collaborate \\ with these different groups? \\ Consider describing any specific tools or processes you use to \\ manage these interactions and ensure smooth operations.\end{tabular} \\ \hline
\textbf{Technology Zero-Shot} & \textbf{Technology Few-Shot} \\ \hline
\begin{tabular}[c]{@{}l@{}}Let's focus on the tools and technologies you use. \\ Can you provide more details about the software you mentioned? \\ What specific features does it have that help you in your tutoring process?\end{tabular} & \begin{tabular}[c]{@{}l@{}}Now, let's focus on the technology aspect: Reflect on your tech use. \\ What specific software or devices do you use for inventory management \\ or communication? Are there any digital tools you wish you had? \\ Consider pain points in your current tech setup and areas where \\ new technology could improve your workflow.\end{tabular} \\ \hline
\end{tabular}%
}
\end{table}

\end{document}